# A unique two-dimensional silver(II) antiferromagnet $Cu[Ag(SO_4)_2]$ and perspectives for its further modifications

Mateusz Domański,*[a] Zoran Mazej,*[b] and Wojciech Grochala*[a]

This work is dedicated to Professor Andrzej W. Sokalski on his $75^{th}$ birthday

[a] Mateusz Domański, Prof. Dr. Wojciech Grochala
Center of New Technologies
University of Warsaw
Zwirki i Wigury 93, 02-089 Warsaw, Poland
E-mail: m.domanski@cent.uw.edu.pl , w.grochala@cent.uw.edu.pl

[b] Dr. Zoran Mazej
Department of Inorganic Chemistry and Technology
Jožef Stefan Institute
Jamova cesta 39, SI-1000 Ljubljana, Slovenia
E-mail: zoran.mazej@ijs.si

Supporting information for this article is given via a link at the end of the document.

**Abstract:** Copper(II) silver(II) sulfate crystallizes in a monoclinic $CuSO_4$-related structure with $P2_1/n$ symmetry. This quasi-ternary compound features $Ag(SO_4)_2^{2-}$ layers, while the remaining cationic sites may be occupied either completely or partially by $Cu^{2+}$ cations, corresponding to the formula of $(Cu_xAg_{1-x})[Ag(SO_4)_2]$, $x = 0.6–1.0$. $CuAg(SO_4)_2$ is antiferromagnetic with large negative Curie-Weiss temperature of –84.1 K and shows two characteristic ordering phenomena at 19 K and 40 K. Density functional theory calculations reveal that the strongest superexchange interaction is a two-dimensional antiferromagnetic coupling within $Ag(SO_4)_2^{2-}$ layers, with the superexchange constant $J_{2D}$ of –11.1 meV. This renders $CuAg(SO_4)_2$ the rare representative of layered $Ag^{2+}$-based antiferromagnets. Magnetic coupling is facilitated by the strong mixing of Ag $d(x^2–y^2)$ and O $2p$ states. Calculations show that $M^{2+}$ sites in $MAg(SO_4)_2$ can be occupied with other similar cations such as $Zn^{2+}$, $Cd^{2+}$, $Ni^{2+}$, $Co^{2+}$, and $Mg^{2+}$.

## Introduction

Two-dimensional (2D) antiferromagnets constitute a very important but rare group of magnetic materials. Their importance stems mainly from the fact that their doping often leads to the appearance of exotic superconductivity. Spin-½ systems with $d^9$ electronic configuration at the metal site are particularly well researched, with $Cu^{2+}$ and $Ni^+$ oxides as key representatives. The magnetic properties of homologous silver(II) compounds have also been extensively studied, with 2D antiferromagnetism, experimentally evidenced[1,2] for binary $AgF_2$. Unfortunately, ternary silver(II) fluorides show collective distortions often resulting in ferromagnetism (e.g. $Cs_2AgF_4$)[3,4]. In some other $Ag^{2+}$ fluorides hole in the *d* band is localized on the $d(z^2)$ orbital[4,5] which reduces magnetic dimensionality to one (1D materials). The only $Ag^{2+}$-based 2D antiferromagnets aside from $AgF_2$ are bis[pentafluoridooxidotungstate(VI)] $Ag(WOF_5)_2$, and two compounds with organic ligands, $Ag(pyz)_2(S_2O_8)$ (pyz = pyrazine) and $Ag(nic)_2$ (nic = nicotinate anion), yet they are rather weakly coupled.[6–9]

In principle, $Ag^{2+}$ oxide systems could present some advantages in this respect as compared to fluorides. As $O^{2-}$ is a more polarizable base than $F^-$, its diffuse valence orbitals may increase electron hopping and thus the coupling. Also, $O^{2-}$ is easier to oxidize than $F^-$, which should increase the strength of the superexchange (SE) mechanism. Unfortunately, the simple binary silver oxide, AgO, has a negative charge-transfer (CT) energy, and it is diamagnetic, so its formula should be written as $Ag_2O·Ag_2O_3$[10]. On the other hand, $Ag_3O_4$ was reported to be paramagnetic above 70 K and the effective magnetic moment per Ag site is 2.03 $\mu_B$[11–13], but a more detailed description of its magnetic properties is still lacking. The silver(III) presence in this compound was confirmed with NMR, although its chemical shift substantially varies from other known $Ag^{3+}$ cations in related oxide systems[14]. Still, it is not clear if $Ag_3O_4$ comprises genuine $Ag^{2+}$ and $Ag^{3+}$ sites as in a localized-electron model ($AgO·Ag_2O_3$), or if it is partly (*e.g.* $Ag_4O_5·Ag_2O_3$) or fully delocalized ($Ag^{2.67+}{}_3O_4$)[12]. Since $Ag_3O_4$ was reported to be metallic[11,12], at least partial delocalization must occur, which indicates that $Ag^{2+}$ can be hole-doped in the oxide environment. In contrast, note that doping has not been achieved so far in 2D fluoride or any other $Ag^{2+}$ systems. The charge delocalization in $Ag_3O_4$ is, in fact, the second known example of hole-doping of the $Ag^{2+}$ system; compounds $AgX·Ag_6O_8$ ($X = NO_3^–$, $ClO_4^–$, $HSO_4^–$) exhibit an average $Ag^{2.67+}$ oxidation state within $Ag_6O_8$ sublattice, and they are also metallic at room temperature (RT) and superconducting at very low temperatures (up to 1.04 K)[15,16].

Complex oxyanions offer more compositional and structural flexibility and stability range than plain oxide ligands. Indeed, a handful of $Ag^{2+}$ salts with oxyanions were reported, *e.g.* silver(II) sulphate[17–19], fluorooxidotungstate(VI)[6], fluorosulfate(VI)[20,21], triflate[22], as well as mixed-valence silver(I/II) fluorosulfate(VI)[21,23] and fluorophosphate[24]. In these oxyanionic salts, the $Ag^{2+}$ sites are connected *via* the oxo ligands which may transfer spin interactions quite effectively. However, most of the mentioned compounds are 1D antiferromagnets: e.g. $AgSO_4$ (–18.7 meV)[17,25,26], $KAg_3(SO_3F)_5$ (–19.1 meV)[23], $Ag_2Ag(SO_3F)_4$ (–15.4 meV)[20,23], or $Ag(SO_3CF_3)_2$ (–9.0 meV)[22]. $Ag^{2+}$ is present here in the square-planar coordination of four oxide ligands, indicating that $d^9$-hole sits on $d(x^2–y^2)$ type orbitals. Aside from these 1D systems, two examples of 2D oxo-ligand magnets are known up to date, *i.e.* antiferromagnetic (AFM) material $Ag[WOF_5]_2$ (–5.4 meV)[6] and ferromagnetic (FM) $Ag(SO_3F)_2$ (+2.9 meV)[20]; the intralayer coupling is noticeably weaker here as compared to the intrachain one in their 1D siblings.

Our recent work showed that a polytype of $Ag^{2+}SO_4$ could also host strong 2D AFM interactions[19]. Unfortunately, this compound was hastily assigned as a "new polymorph" of $AgSO_4$ before we discovered the crucial role which $Cu^{2+}$ impurity plays in the formation of this phase. We have now been able to resolve the true nature of these samples, and we present here a revised crystal structure and properties of heterocationic quasi-ternary sulfate with the formula $(Cu_xAg_{1–x})[Ag(SO_4)_2]$, $x$ = 0.6–1.0 (for x=1 this corresponds to $CuAg(SO_4)_2$). Based on experimental results and density functional theory (DFT) calculations we show that this compound is a new rare example of 2D AFM and also the first member of the $MAg(SO_4)_2$ prospective family of such systems.

## Results and Discussion

### Crystal structure of $Cu^{II}Ag^{II}(SO_4)_2$

Since the compound[19] previously described as silver(II) sulfate, was now found to contain up to 15% (molar) copper, *cf.* **Supporting Information** (**SI**, **Figure S1**), we have studied more compositions across the $AgSO_4$/$CuSO_4$ phase diagram. We have employed an acid displacement reaction, where weaker and volatile acid HF is eliminated by anhydrous $H_2SO_4$ poured onto a mixture of $CuF_2$ and $AgF_2$, with the reaction vessel cooled in liquid-nitrogen (**Figure S2**):

$$AgF_{2(s)} + CuF_{2(s)} + 2\ H_2SO_{4(l)} \rightarrow CuAg(SO_4)_{2(s)} + 4\ HF_{(g)}\uparrow$$

The remaining $H_2SO_4$ was washed off multiple times with large portions of anhydrous HF.

For an $AgF_2$:$CuF_2$ reagent ratio of 1:1, we have obtained brown, heterocationic sulfate $CuAg(SO_4)_2$ (**Figure S3**). The powder XRD pattern of the reaction product is shown in **Figure 1** together with the diffraction profile calculated using the Rietveld method[27]. The experimental profile is well described by a single crystalline phase. The crystal structure was refined using GSAS-II software[28]. The $CuAg(SO_4)_2$ crystallizes in the monoclinic system, *cf.* **Table 1**. We chose the non-standard setting $P2_1/n$ for the No. 14 space group to keep a similar orientation as in the related $CuSO_4$ structure[29]. Further details of the crystal structure reported here may be obtained from CCDC/FIZ by quoting CSD deposition No. 2262020. Crystallographic Information File is also provided in the **SI**.

The crystal structure of $Cu^{II}Ag^{II}(SO_4)_2$ is analogous to that of a recently discovered mineral $Mg^{II}Cu^{II}(SO_4)_2$[30] (**Figure 2**) while both are $P2_1/n$ monoclinic distortions of the orthorhombic $Cu^{II}SO_4$. Unlike $CuSO_4$, which is composed of Cu-O-Cu chains along *b* lattice vector, both monoclinic quasi-ternary structures comprise the $M(SO_4)_2^{2–}$ sublattices (M = $Ag^{2+}$ for $CuAg(SO_4)_2$ or $Cu^{2+}$ for $MgCu(SO_4)_2$) parallel to (–101) crystallographic planes (**Fig. S4**). XRD structure determination indicates that the volume of mixing for the formation of $CuAg(SO_4)_2$ is slightly positive, *i.e.* it reaches +2.3% of the combined volume of two separate metal sulfates, which agrees well with DFT calculations (**Table S1**).

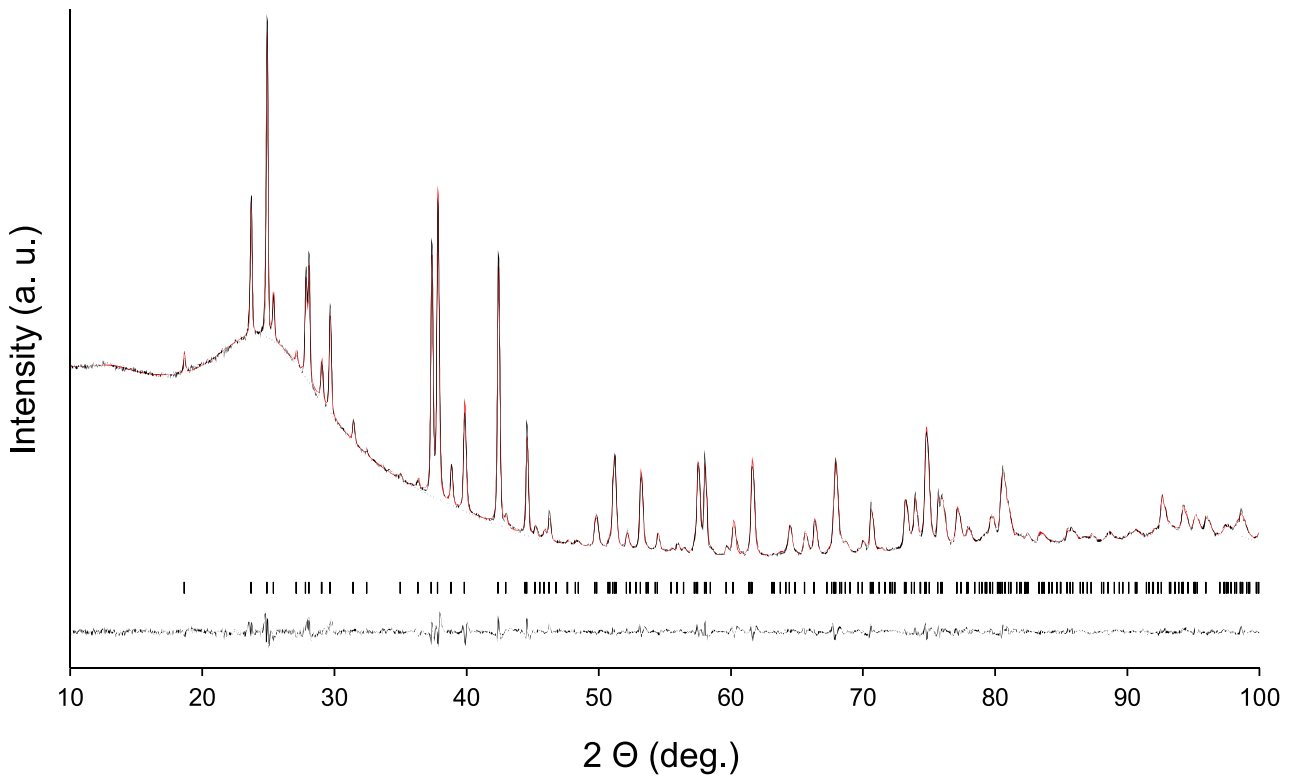

**Figure 1.** The powder XRD pattern of $CuAg(SO_4)_2$ (black), the refined pattern (red), and the fitted polynomial background (dashed line). The markers for $CuAg(SO_4)_2$ phase for Co Kα₁ wavelength and differential profile are below.

**Table 1.** Crystallographic data for $CuAg(SO_4)_2$ crystal structure.

| Parameters | $CuAg(SO_4)_2$ | | |
|---|---|---|---|
| Crystal system, space group | Monoclinic | $P2_1/n$ (No. 14) | |
| Formula units, volume [Å$^3$]: | Z = 4 | 294.67(1) | |
| a, b, c [Å]: | 4.73365(8) | 8.71928(11) | 7.15754(16) |
| α, β, γ [°] | 90 | 94.0852(9) | 90 |
| Radiation type, 2Θ range | Co Kα | 10° < 2Θ < 100° | |
| Fit parameters | $R_p$ = 1.24% | $wR_p$ = 1.80% | GOF = 1.77 |

The structural similarity of $CuAg(SO_4)_2$, $MgCu(SO_4)_2$, and parent $CuSO_4$, as well as the chemical similarity of $Cu^{2+}$ and $Ag^{2+}$ naturally suggests a possibility that $CuAg(SO_4)_2$ could be non-stoichiometric, *i.e.* a solid solution of $Cu^{2+}$/$Ag^{2+}$ cations at both metal sites. For this reason, the Rietveld refinements allowing for diverse occupation at metal sites were also conducted for the sample obtained from $AgF_2$:$CuF_2$ reagent ratio of 1:1. However, these refinements did not improve pattern fit parameters, thus suggesting that each crystallographic site is occupied by either Cu or Ag but not both cations (*i.e.* it is ordered rather than disordered structure). We note that the M-O bond lengths for the Cu site equal 1.954(4), 1.965(5), and 2.373(7) Å, while for the Ag site, they equal 2.086(4), 2.130(6), and 2.788(6) Å. Considerably different distances make Ag ↔ Cu replacement difficult.

The ordered model has been further supported by DFT+U modelling. We have tested three structural models for 1:1 $CuAg(SO_4)_2$ originating from various Ag/Cu permutations at four cationic sites in the unit cell. The model originating from the Rietveld fit had the lowest energy, while one with a single-site substitution and with fully inversed Ag ↔ Cu occupations on both sites had energy higher by +25.0 kJ/mol and +45.5 kJ/mol respectively. This reconfirms that both cations occupy their preferred distinct crystallographic sites in the 1:1 compound.

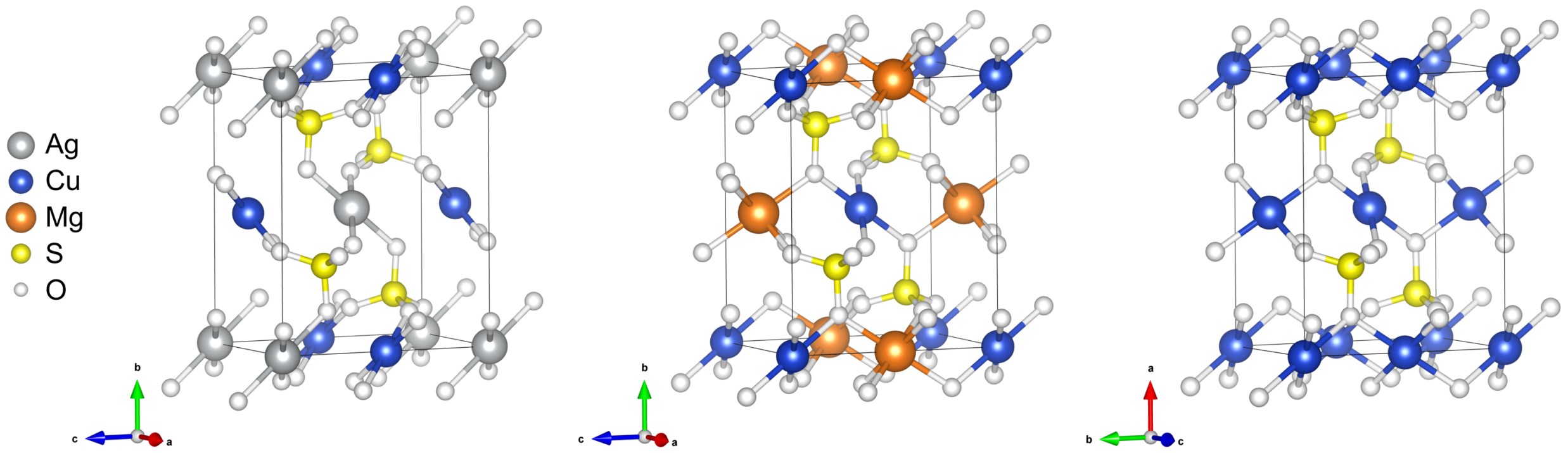

**Figure 2.** The $CuAg(SO_4)_2$ crystal structure (left) is analogous to another quasi-ternary sulfate $MgCu(SO_4)_2$ (middle) while both are related to the orthorhombic $CuSO_4$ (right). Chains along *c* in the two monoclinic quasi-ternaries are heterocationic and they lack a covalent O'-M-O" bridge as present in $CuSO_4$ along *b*.

As to non-stoichiometric compositions, a possibility arises that partial Cu → Ag substitution could occur solely at the Cu site. To test this hypothesis, we prepared samples with smaller Cu concentrations at initial 15:85 and 5:95 Cu:Ag molar ratios. These reactions resulted in changes in the molar volume of the monoclinic phase up to +1.85%, suggesting the replacement of some copper for silver (**Figures S5-7**, and **Table S2**). Moreover, in each of these samples, partial Cu → Ag substitution on the Cu site led to an improvement of the profile fit R factors. This implies that only Cu → Ag substitution at the Cu site of $CuAg(SO_4)_2$ structure occurs, resulting in $Cu_xAg_{1-x}(SO_4)_2$ with $x < 0.5$. Importantly, no evidence for Ag → Cu substitution at the Ag site was found. Interestingly, in samples obtained from Cu:Ag fluorides of 15:85 and 5:95 ratios, we have not observed changes in $AgSO_4$ *C*2/*c* phase[26] where unit cell volume agrees with the known one within uncertainties. This suggests that the *C*2/*c* phase does not have the flexibility exhibited by the $P2_1/n$ one. Based on all attempts conducted, we could deduce that the overall Cu concentration in the $P2_1/n$ phase is in ~30% ≤ 50% range.

To further support the hypothesis of a partial Cu → Ag substitution, we have performed DFT+U calculations for the entire composition range (**Figures S8** and **S9**). Primarily, for a 1:1 ratio, *i.e.* $Cu_{0.5}Ag_{0.5}SO_4$ we predict that the formation energy is the most negative, yielding –4.0 kJ per mole of $MSO_4$ (–3.7 kJ per mole at HSE06 level, *cf.* **Table S3**). The situation in the sulfate system thus differs from the case of 1:1 ternary fluoride, where $Cu_{0.5}Ag_{0.5}F_2$ was predicted to be energetically not favored (+3.1 kJ/mol per $MF_2$ with the HSE06 method)[31]. This explains why the 1:1 system ($Cu_{0.5}Ag_{0.5}SO_4$) does not undergo phase separation into two distinct Cu-rich and Ag-rich phases, in contrast to related $Cu_{0.5}Ag_{0.5}F_2$[32].

Moreover, applying the convex hull approach to the DFT+U obtained phase diagram leads to a conclusion that the energetically stable quasi-ternary phases should have a composition of $0.25 < x < 0.5$ for $Cu_xAg_{1-x}SO_4$ formula; this is indeed in agreement with observed Cu contents in our samples. For samples exceeding this composition range, the system should decompose to quasi-ternary and quasi-binary sulfates, which is consistent with the experimental results (*i.e.* obtaining both *C*2/*c* and $P2_1/n$ phases for a 15% Cu sample synthesis).

Finally, it turns out that $Cu_xAg_{1-x}SO_4$ in the $P2_1/n$ crystal structure is more energetically stable than binary substrates (in their experimentally known forms) for the entire composition range $0 \leq x \leq 1$. This is an interesting result for Cu-rich compositions, as it suggests that even pure $CuSO_4$ could be more stable in the monoclinic $P2_1/n$ rather than in orthorhombic form (by –1.1 kJ/mol). The same calculation using HSE06 yields a qualitatively similar result (–2.2 kJ/mol) but SCAN indicates energy near-equivalence of both forms (+0.2 kJ/mol). The preference for monoclinic polymorph of $CuSO_4$ at low temperatures remains an open question[33] (**Table S3**).

Further confirmation of $CuAg(SO_4)_2$ structural distinctness comes from vibrational spectroscopy measurements, as shown in **Figure S10** and **Table S4** compared with $AgSO_4$ and $CuSO_4$. Vibrational spectroscopy has been routinely applied as a method of characterization of $Ag^{2+}$ compounds[17–19,6,26,34–37]. Since the specimen was single-phase, all of the observed bands can be assigned to the title compound. In particular, the Raman scattering pattern for $CuAg(SO_4)_2$ contrasts significantly from those of individual metal sulfates, confirming its different symmetry and proving that Raman spectroscopy can be used for identification of the new compound.

## Magnetic properties of $Cu^{II}Ag^{II}(SO_4)_2$

$CuAg(SO_4)_2$ is a heterocationic ordered crystal, with the cations from different rows of the periodic table and both having an open shell $d^9$ spin–½ configuration. In the vast majority of such 3d-4d compounds either one cation is diamagnetic or there is a substitutional disorder in the crystal structure, hence magnetic properties of $Cu^{II}Ag^{II}(SO_4)_2$ are of interest.

The magnetic susceptibility $\chi(T)$ of $CuAg(SO_4)_2$ between 2 and 300 K in the zero-field-cooled (ZFC) and field-cooled (FC) regimes is shown in **Figure 3**. The shape of magnetic susceptibility dependence below 45 K indicates a spontaneous magnetic ordering in the sample, appearing as weak ferromagnetism. The transition temperature ($T_t$) determined from the maximum of $-d\chi/dT$ is $T_{t1}$ = 40.2 K. However, there is a second broader maximum at about $T_{t2}$ = 19.0 K. Above 45 K the sample shows paramagnetic behavior, following Curie–Weiss law, $\chi = C/(T - \theta)$, fairly well (**Figure 4**). The Curie constant may be estimated at $C$ = 0.419 emu·K/mol, which agrees with the expected value for the spin-only $S$ = ½ $Cu^{2+}$ and $Ag^{2+}$ ions. The effective magnetic moment, $\mu_{eff}$, averaged for all paramagnetic centers in $CuAg(SO_4)_2$, is 1.83 $\mu_B$. The magnetic behavior of $CuAg(SO_4)_2$ is strikingly similar to that of $Ag(WOF_5)_2$[6] and $Cu_2V_2O_7$[38] which also display weak ferromagnetism. Measured here $\mu_{eff}$ is close to the average $\mu_{eff}$ for $Ag^{2+}$ (1.7 $\mu_B$) and $Cu^{2+}$ (1.9 $\mu_B$) in these two compounds (summary in **Table 2**). Moreover, $CuAg(SO_4)_2$ has a negative paramagnetic Curie–Weiss temperature, $\theta = -84.1$ K, just like $Ag(WOF_5)_2$ and $Cu_2V_2O_7$ (−63 K and −66 K, respectively), suggesting the presence of predominant AFM interactions in all these systems. As we will see, these interactions may be associated with the presence of the $[Ag(SO_4)_2]^{2-}$ sublattice.

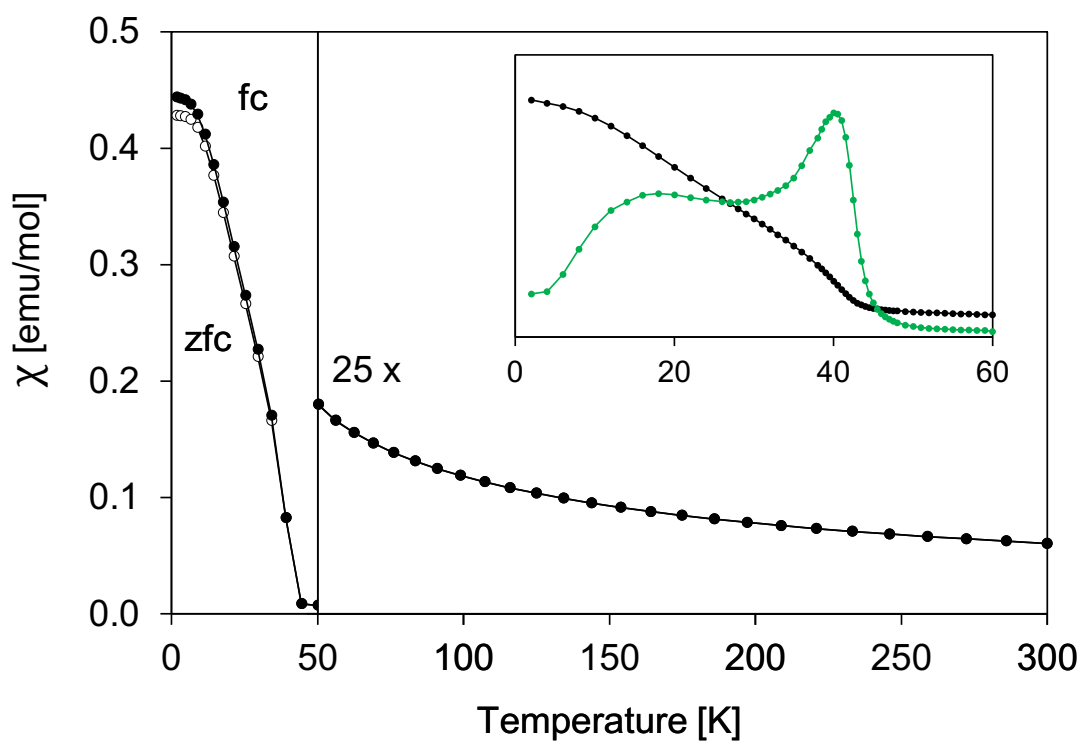

**Figure 3.** Temperature-dependent magnetic susceptibility $\chi$ measured in a magnetic field of 1 kOe. The inset shows magnification on magnetic susceptibility (black) and $-d\chi/dT$ (green) for the 2-60 K temperature range measured in the FC regime applying a 10 kOe magnetic field.

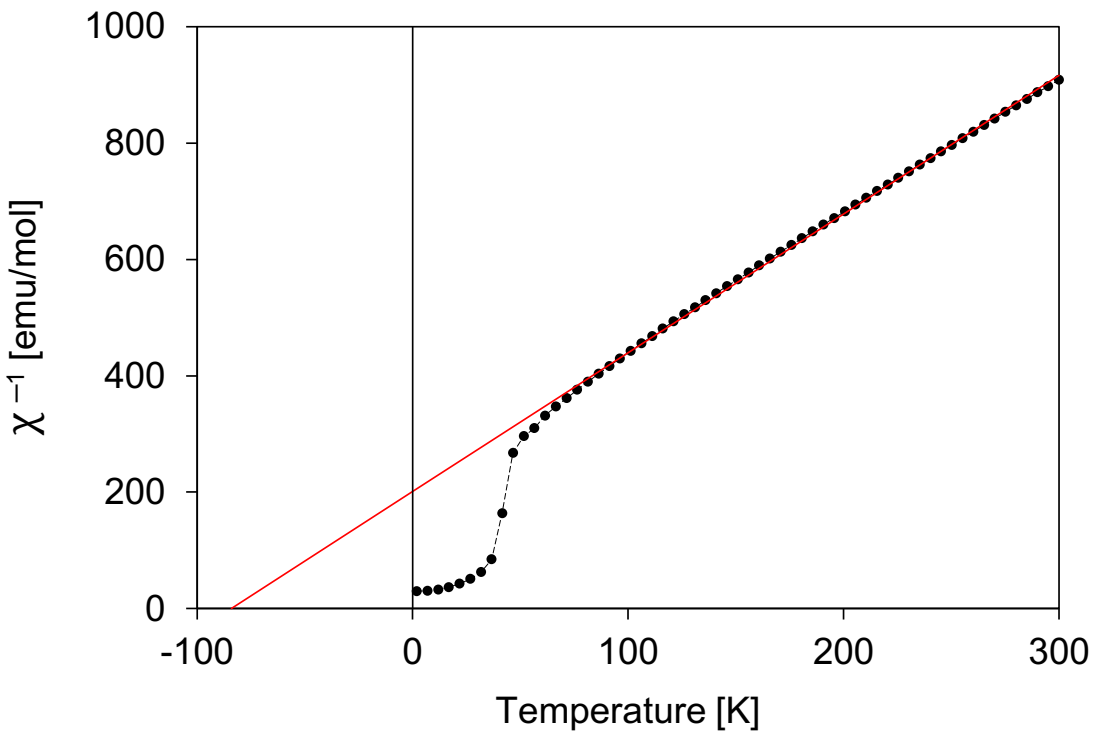

**Figure 4.** Temperature dependence of the reversed magnetic susceptibility $\chi$ measured in a magnetic field of 10 kOe in the FC regime, showing a good correlation with the Curie-Weiss law (red line, $R^2$=0.999).

To obtain an improved insight into the weak FM response of $CuAg(SO_4)_2$ we measured the isothermal magnetization curves above $T_{t1}$, between $T_{t1}$ and $T_{t2}$, and below the $T_{t2}$. The $M(H)$ measured at three different temperatures, is presented in **Figure 5**. At 300 K the magnetization changes linearly with the magnetic field, with remnant magnetization of about $5 \cdot 10^{-5}$ $\mu_B$ and barely any coercivity. However, below the $T_{t1}$ temperature, *i.e.* at 4 K and 35 K, the magnetization curves reveal the presence of hysteresis loops, with a remnant magnetization of approximately 0.039 $\mu_B$ per "averaged" metal cation and a coercive field of $H_c = 238$ Oe at 4 K. The maximum magnetization in the largest magnetic field applied of 40 kOe at 4 K is only 0.10 $\mu_B$ per averaged $M^{2+}$, so it is not saturated at the experiment conditions and shows a nearly linear increase with the $H$ field. The expected saturation magnetization equals $S \cdot g \cdot \mu_B \approx 1$ $\mu_B$/cation for both $S = ½$ spin cations (*e.g.* as for genuinely FM $Ag(SO_3F)_2$), while the value observed for $CuAg(SO_4)_2$ is much smaller. This suggests that the net FM moment comes from the mutual AFM alignment of $Cu^{2+}$ and $Ag^{2+}$ sublattices (with slightly different moments in each sublattice) and/or from spin canting. Considering that $CuAg(SO_4)_2$ displays similar behavior as $Ag(WOF_5)_2$ and $Cu_2V_2O_7$, the latter scenario is very likely.

**Table 2.** Comparison of $CuAg(SO_4)_2$ magnetic properties with similar $Cu^{2+}$ and $Ag^{2+}$ systems. Data include magnetic transition temperature, Curie-Weiss temperature, effective magnetic moment, the predominant exchange coupling constant, and maximum magnetic moment at the given experimental conditions.

| System | $T_t$ (K) [a] | Θ (K) | $\mu_{eff}$ ($\mu_B$) | $J_{2D}$ (meV) | $\mu_{max}$ ($\mu_B$) |
|---|---|---|---|---|---|
| $CuAg(SO_4)_2$ | 40.5 / 19.0 | −84.1 | 1.83 [b] | −11.1 [c] | 0.10 [b] (40 kOe) |
| $Cu_2V_2O_7$ | 35 [38] | −66 | 1.9 | ND | 0.04 (5 kOe) |
| $CuSO_4$ | 34.5 [39] | −77.5 | ND | −5.5 [19]c | ND |
| $CuF_2$ | 69 [40] | −200 | ND | −17 / −17.4 [41]c | ND |
| $Ag(nic)_2$ | 11.8 [8] | −46 | ND | −2.6 | ND |
| $Ag(pyz)_2(S_2O_8)$ | 7.8 [7] | −66.8 | ND | −4.6 | ND |
| $Ag(WOF_5)_2$ | 20.7 [6] | −63 | 1.7 | −5.4 | 0.07 (50 kOe) |
| $Ag(SO_3F)_2$ | 24.8 [20] | +34.1 | 1.94 | +2.9 | 1.08 (50 kOe) |
| $AgF_2$ | 163 [1] | −715 | 2.0 | −70 / −55.8 [42]c | 0.01 (ND) |

[a] Néel temperature of a 2D system is strongly affected by the interlayer coupling strength [31,32]. [b] average over $Ag^{2+}$ and $Cu^{2+}$ cations. [c] values calculated with HSE06 functional. ND – not determined.

The questions remain what are the reasons for the observed weak ferromagnetism and to which particular phenomenon could be the two $T_t$ assigned. The weak ferromagnetism in such systems is often due to spin canting caused by Dzyaloshinsky-Moriya interaction[43,44] that should strongly depend on the ligand nature and details of the crystal structure. Comparison of $Cu^{2+}$ and $Ag^{2+}$ compounds with similar structures and the same ligands, $CuF_2$ and $AgF_2$, reveal that the higher $T_t$ is observed for the latter, due to both the stronger intra-sheet and inter-sheet SE.[2,40–42] Also, a larger canting angle has been reported for $Ag^{2+}$ (0.5°)[1] in $AgF_2$ than $Cu^{2+}$ in $CuF_2$ (0.1°)[40]. This is due to a much stronger spin-orbit coupling for $Ag^{2+}$ than for $Cu^{2+}$, as the roots of this effect are relativistic, and its strength increases with atomic number. Finally, a stronger antisymmetric exchange is expected as the exchange interactions between the spins are weaker. All this renders analysis very complicated. Ultimate assignment of two ordering events observed in magnetometry should be possible in the future with the use of muon spin resonance spectroscopy and after confirmation of their intrinsic character.

Magnetic properties of $CuAg(SO_4)_2$ and similar systems are presented in **Table 2**. Overall, $AgF_2$ and the discussed oxyanionic $Ag^{2+}$ 2D systems share a similar structural motif (**Figure 6**) *i.e.* an $AgL_{4/2}$ 2D network (L – ligand). $Ag(SO_3F)_2$ is the only one in this family with a positive Curie-Weiss temperature, indicating that FM interactions prevail in this compound. According to the Goodenough-Kanamori-Anderson rules[45–47], this may be associated with the smallest Ag'-L-Ag" angle of 127.6° in this system as calculated for central atom S. The respective angle in $CuAg(SO_4)_2$ is more open, some 134.2°, and this drives the leading intra-sheet interaction from FM towards AFM character. An additional complication is that – as suggested earlier[20] – the SE pathway may omit the S atom, and involve Ag'-O···O-Ag'' bridge with O atoms not connected by a proper chemical bond. Such a mechanism known as super-superexchange (SSE) has been already discussed in both sulfates and phosphates[48–50]. In the fluorosulfate, the two Ag-O···O angles are 161.6° and 114.3, with the corresponding values for $CuAg(SO_4)_2$ of 157.5° and 135.0°. The smaller of these two angles is much closer to 90° for the fluorosulfate than for $CuAg(SO_4)_2$, which may contribute to the different magnetic properties of these related compounds.

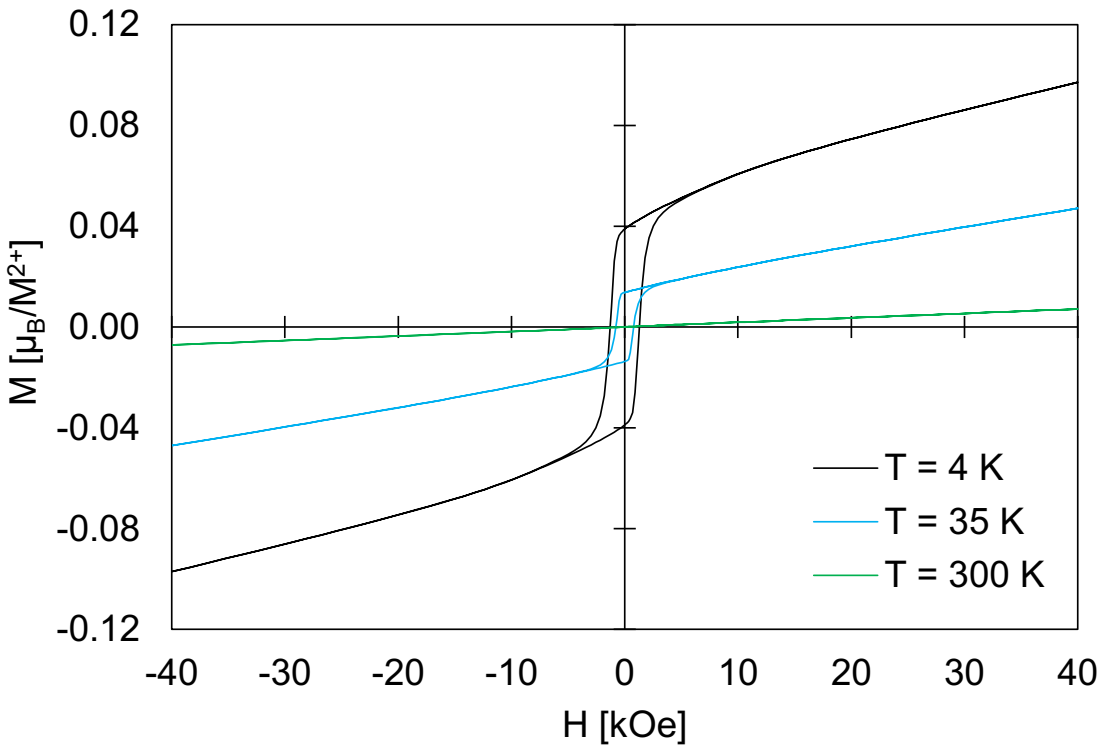

**Figure 5.** The magnetic moment M(H) for $CuAg(SO_4)_2$ at 4 K, 35 K, and 300 K.

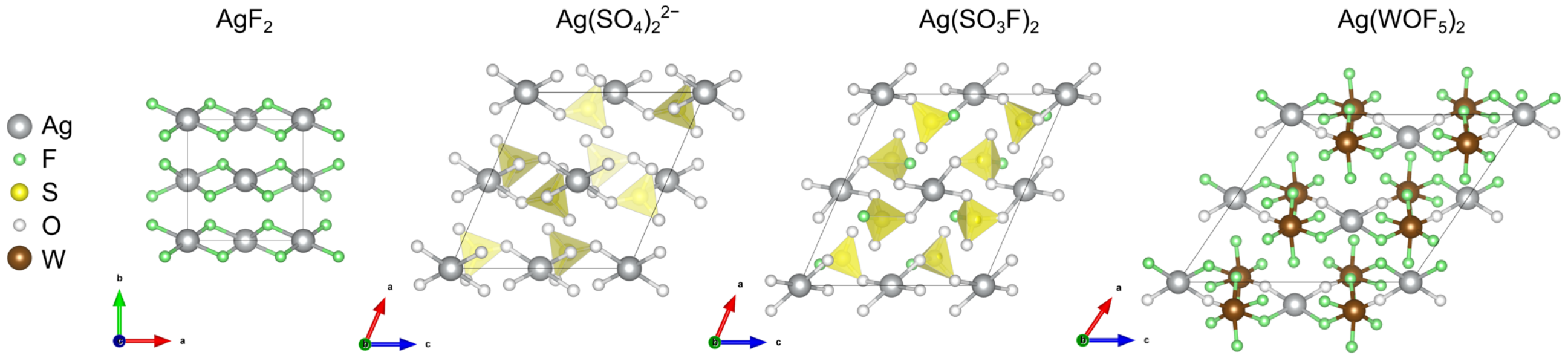

**Figure 6.** The crystal structure of silver(II) compounds exhibiting 2D magnetism. The Ag fractional coordinates in $CuAg(SO_4)_2$ and fluorosulfate are the same as in $AgF_2$ despite the monoclinic distortion of the unit cell, while in $Ag(WOF_5)_2$ the cations at *c* = 0.5 are shifted, and cause buckling of the layers.

## Superexchange interactions and electronic structure

**Table 3.** The exchange interactions geometric parameters and energy in the $CuAg(SO_4)_2$ crystal structure, as calculated using HSE06.

| Interaction | M'-O [Å] | O-M" [Å] | M'-M" [Å] | ∠SE M'-S-M" | ∠SSE M'-O···O | ∠SSE M"-O···O | E [meV] |
|---|---|---|---|---|---|---|---|
| $J_1^{(Cu\text{-}Ag)}$ [001] | 1.933 (Cu) | 2.106 (Ag) | 3.582 | 66.8° | 106.0° | 106.5° | +1.2 |
| $J_2^{(Ag\text{-}Ag)}$ [100] | 2.092 | >2.8 [a] | 4.732 | - | - | - | –0.2 |
| $J_2^{(Cu\text{-}Cu)}$ [100] | 1.933 | 1.949 | 4.732 | 96.4° | 144.6° | 102.2° | +0.6 |
| $J_3^{(Cu\text{-}Ag)}$ (001) | 1.949 (Cu)<br>1.933 (Cu) | 2.106 (Ag) | 4.949 | 102.0°<br>102.0° | 148.0°<br>142.9° | 111.9°<br>96.9° | –0.2 |
| $J_{1D}^{(Cu\text{-}Ag)}$ [201] | 1.949 (Cu) | 2.092 (Ag) | 5.711 | 123.0° | 138.9° | 153.4° | –10.3 |
| $J_{2D}^{(Ag\text{-}Ag)}$ (-101) | 2.092 | 2.106 | 6.002 | 134.0° | 135.1° | 158.8° | –11.1 |
| $J_{2D}^{(Cu\text{-}Cu)}$ (-101) | 1.949 | >2.4 [a] | 6.002 | - | - | - | +0.4 |

[a] The half-occupied $d(x^2–y^2)$ orbital is orthogonal to these bonds.

To gain further insight into the magnetic properties of the title compound, we have conducted DFT calculations of the underlying magnetic interactions. Here, the interactions in $CuAg(SO_4)_2$ along possible SE paths and within ~6.0 Å were considered, all of which are shown in **Figures S11** and **S12**. The exchange coupling constants were calculated using the Heisenberg Hamiltonian of the type $\boldsymbol{H} = -\sum_{\langle ij\rangle} \boldsymbol{J_{ij} s_i s_j}$ (full version in **Equation S1**), based on the energies calculated for different spin states. Geometries of SE and SSE pathways together with calculated exchange constants are summarized in **Table 3**.

According to our HSE06 results, there are three main components of exchange interactions in $CuAg(SO_4)_2$. The strongest one is 2D AFM interaction $J_{2D}^{(Ag\text{-}Ag)}$ (or simply $J_{2D}$) parallel to (–101) crystallographic plane of the unit cell, with a constant equal –11.1 meV (the hosting $[Ag(SO_4)_2]^{2–}$ layers are shown in **Figure 7**). This value is noticeably smaller than experimentally determined for 1D AFM, $AgSO_4$, being –18.7 meV;[17,26] the difference may arise from smaller Ag-S-Ag angles in $CuAg(SO_4)_2$ than in *C*2/c $AgSO_4$ (134.2° *vs.* 165.9° on average). Comparable in size to $J_{2D}$ is the one-dimensional interaction $J_{1D}^{(Cu\text{-}Ag)}$ between the two 3d and 4d cations along [201] diagonal with a SE constant of –10.3 meV. There is also a weak FM interaction $J_1$ along [001] direction, which also involves $Cu^{2+}$ and $Ag^{2+}$. However, $J_1$ strength varies strongly with the functional used +1.0 (DFT+U), +1.2 (HSE06), up to +7.3 meV (SCAN). The full comparison of results using these three DFT methods is available in **Tables S5** and **S6**, yet except for $J_1$ these results agree qualitatively in all other calculated exchange constants. The geometry of the [001] chain with 66.8° Cu-S-Ag angle alone, suggests that the $J_1$ exchange would be FM, and both cations have their $d(x^2–y^2)$ orbitals directed towards mediating $(SO_4)^{2–}$ groups. Despite $d(x^2–y^2)$ orbitals of the atoms involved in $J_1$ being aligned almost in parallel along [001] direction, the distance between Ag and Cu along *c* is only 3.58 Å so a weak direct exchange might also contribute to $J_1$.

We notice also several other weaker SE interactions in the crystal structure of $CuAg(SO_4)_2$. Cations of $Ag^{2+}$ should not interact significantly along [100] with $J_2^{(Ag\text{-}Ag)}$ because of a lack of a direct Ag-$SO_4$-Ag bridge along this direction. On the other hand, each $Cu^{2+}$ cation forms two Cu'-$SO_4$-Cu" bridges, resulting in weak FM interaction with constant $J_2^{(Cu\text{-}Cu)}$. Such bridges are common in 11th group oxyanionic salts, e.g. the above-mentioned silver(II) triflate (with an experimental $J_{1D} = -9.0$ meV), but we estimate that this interaction would be weak in $CuAg(SO_4)_2$. Another interesting motif is formed by layers parallel to (001) with each $Cu^{2+}$ interacting *via* four $SO_4^{2-}$, while each $Ag^{2+}$ interacts through two ligands. Despite both Ag-S-Cu and SSE angles close to 90°, the $J_3^{(Cu-Ag)}$ interaction is weakly AFM. Considering the whole (001) plane one can notice a quasi-hexagonal lattice of cations involving both $J_2^{(Ag-Ag)}$, $J_2^{(Cu-Cu)}$ interactions and $J_3^{(Cu-Ag)}$. By forming three different interactions the spin lattice escapes the frustration.

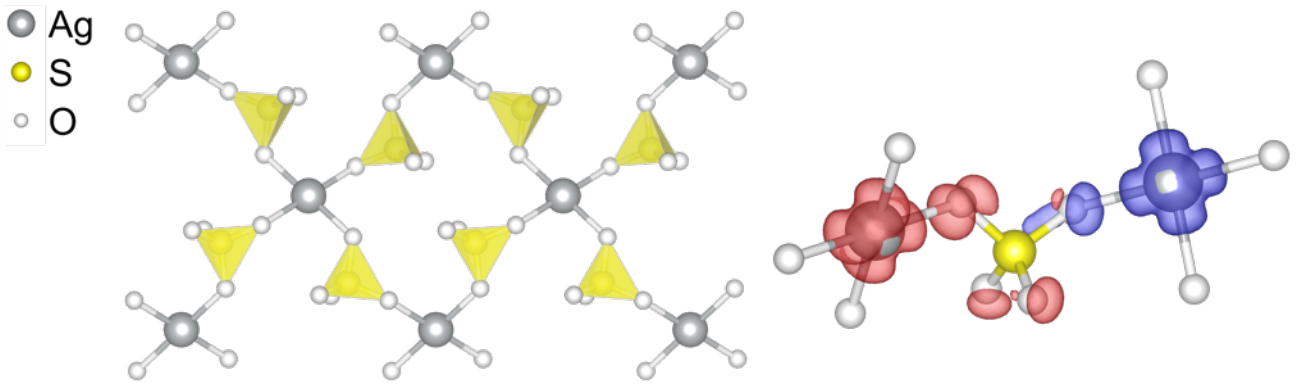

**Figure 7.** The $Ag(SO_4)_2^{2-}$ layer from $CuAg(SO_4)_2$ and a close-up, showing the spin-density along the interaction pathway measured by $J_{2D}$ (red for spin-up, blue for spin-down). Electrons in a sulfate anion mediate SE interaction due to high S-O bond covalence. Notice that net spin on sulfate anion is not null, indicating its quasi-radical character. The spin-density calculated using HSE06 functional is shown with its isosurface value set at 0.005 e/Å$^3$.

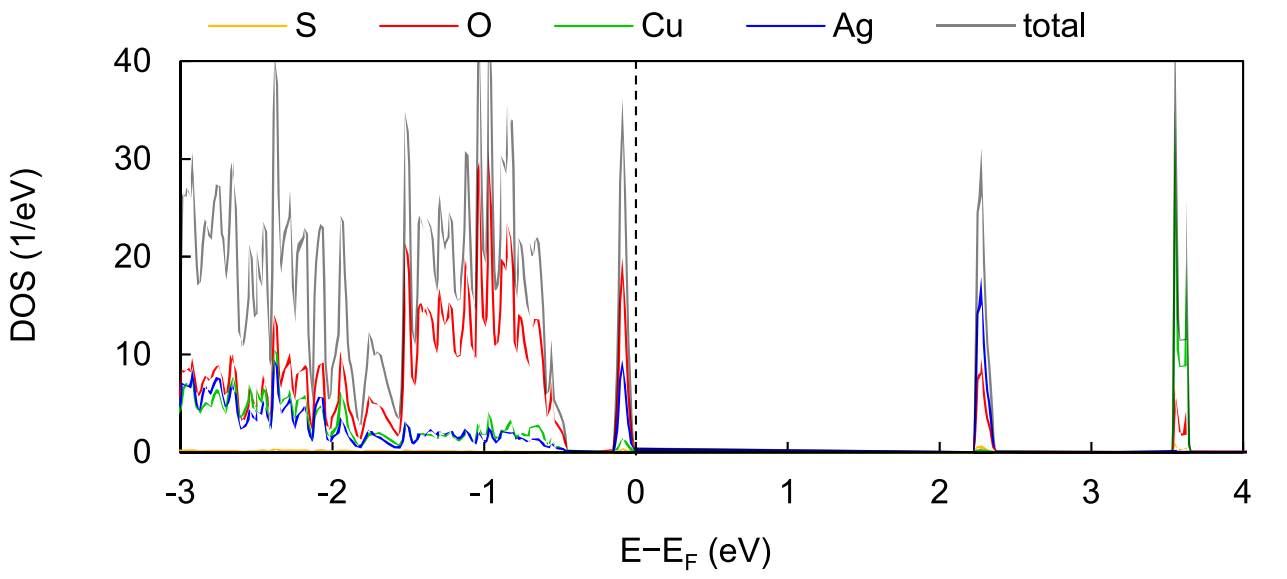

**Figure 8.** Total and partial electronic density of states (DOS) projected on S, O, Cu, and Ag atoms for $CuAg(SO_4)_2$ structure calculated with HSE06 functional. The Fermi energy is indicated with the dashed line.

The DFT results indicate that magnetic interactions in $CuAg(SO_4)_2$ are rather complex. The two strongest interactions $J_{1D}$ and $J_{2D}$ (see **Figure S13**, **Table S6**) are not frustrated in the ground state and intersect at an angle of 20.9° leading to a complex 3D magnetic lattice. Their presence should account for the experimentally determined antiferromagnetism, as reflected by the negative value of $\theta = -84.1$ K. However, analytical functions for magnetic susceptibility typical of spin-½ square lattice[51] or 1D chain[51,52] interacting through Heisenberg exchange could not be fitted to the experimental data.

The exchange interactions between $Ag^{2+}$ centers are mediated by sulfate dianions, which show partial radical character caused by oxidizing properties of $Ag^{2+}$. An uncompensated magnetization of each $SO_4^{2-}$ reaches ±0.113 $\mu_B$, displayed as spin-density isosurface around the ligand in **Figure 7**. Such strong spin-polarization of sulfate groups reduces magnetization on $Ag^{2+}$ cations to ±0.567 $\mu_B$. With two $SO_4^{2-}$ per $Ag^{2+}$ the net magnetization of $Ag(SO_4)_2^{2-}$ unit is ±0.793 $\mu_B$, equal to that of $Cu^{2+}$ cation.

The $Ag^{2+}$ oxidizing character implies also that the valence Ag(d) and O(p) orbitals must strongly mix. Indeed, the electronic density of states (DOS) calculated for $CuAg(SO_4)_2$, **Figure 8**, shows that $Ag^{2+}$ $d(x^2-y^2)$ bands and oxide $p$ bands form a separate pocket of states just below Fermi energy. The valence band is predominated by O states, though Ag contribution is also substantial indicating its CT character. The conduction band corresponds to the upper-Hubbard band of the $Ag^{2+}$ $e_g$ hole, while the $Cu^{2+}$ $e_g$ hole is located at about 1.3 eV higher energy. $CuAg(SO_4)_2$ should thus be classified as a CT gap semiconductor according to the Zaanen-Sawatzky-Allen classification[53], its frontier orbitals originating from the $Ag(SO_4)_2^{2-}$ sublattice. The computed band gap is relatively broad, some 2.23 eV at the hybrid DFT level of theory, though smaller as compared to 2.51 eV for $AgF_2$ (calculated with the same method)[31] (*cf.* **Table S3 in ESI**).

## Perspectives for $MAg(SO_4)_2$ family

The similarity between $MgCu(SO_4)_2$ and $CuAg(SO_4)_2$, together with the observed partial Cu → Ag substitution in the latter, suggests that the Cu site in the $P2_1/n$ structure of $CuAg(SO_4)_2$ could undergo a substitution. We have tested this surmise by fully optimizing the crystal structures of several $M^{2+}[Ag(SO_4)_2]$ systems. Here, $M^{2+}$ includes closed-shell cations from the 2$^{nd}$ or 12$^{th}$ groups of the Periodic Table and two open-shell cations from the 3d-block which may resist oxidation by $Ag^{2+}$(**Table 4**).

According to our calculations with HSE06 functional, several $MAg(SO_4)_2$ derivatives with small cations are stable with respect to phase separation to quasi-binary sulfates (**Figure 9**). We note a relation between the effective crystal radii ($R_C$) corresponding to octahedral coordination (CN = 6)[54], and the energy of quasi-ternary sulfate formation (ΔE) in the reaction: $AgSO_4 + MSO_4 \rightarrow MAg(SO_4)_2$. Generally, the smaller the $M^{2+}$ cation the more negative the energy of formation. This may be easily understood as the parent $P2_1/n$ structure features a small octahedrally coordinated Cu site. Thus, for much larger $M^{2+}$ cations such as $Ba^{2+}$, the structure must deform to provide more M-L bonds (*i.e.* to achieve CN > 6, typical for such cations), and that decreases its stability. Together with increasing $M^{2+}$ cation radius, we notice an increase of the Ag'-S-Ag" angle, *i.e.* flattening of $[Ag(SO_4)_2]^{2-}$ layers, showing their flexibility (**Figure S14**, **Table S7**).

Also, we have calculated $J_{2D}$ coupling constants in these hypothetical $MAg(SO_4)_2$, as shown in **Figure 9**. We find a strong correlation between $J_{2D}$ and Ag'-S-Ag'' angle (α), as expected based on Goodenough-Kanamori rules. A potential limit of the antiferromagnetic interaction strength within the $[Ag(SO_4)_2]^{2-}$ layers is calculated for the $Ba^{2+}$ derivative ($J_{2D}$ of –50 meV); this is close to the value for $AgF_2$ of –56 meV obtained with the same method[2].

The possibility of the 2D $MAg(SO_4)_2$ lattice formation with various $M^{2+}$ cations is a considerable functional advantage of the sulfate system as compared to known fluoride or oxide systems. The fluorides tend to lose 2D AFM character upon formation of ternaries, while there are no ternary oxides systems at all containing $Ag^{2+}$. On the contrary, $MAg(SO_4)_2$ stoichiometry offers the possibility to prepare a 2D AFM system with diamagnetic $M^{2+}$ cations and thus with a much simpler magnetic structure than that of the discussed $CuAg(SO_4)_2$ derivative. Hypothetical substitutions of $Cu^{2+}$ for different valence cations, or sulfate dianions for isostructural $PO_3F^{2-}$ or $SO_3F^{-}$ could also be considered in the future[55].

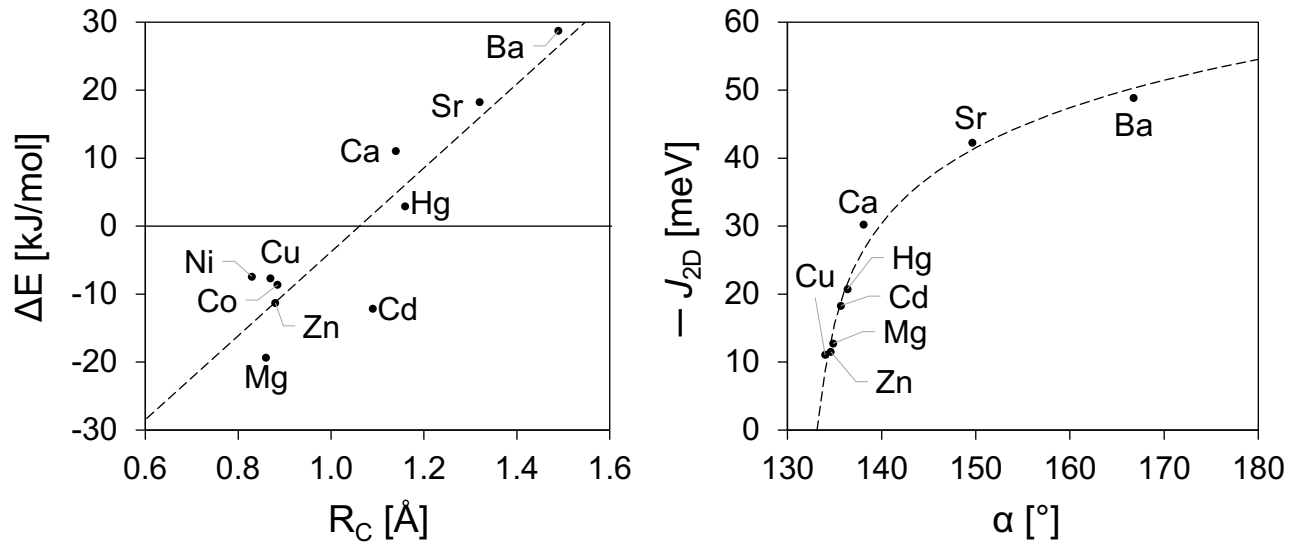

**Figure 9.** (Left) The formation energy (ΔE) of several $MAg(SO_4)_2$ compounds versus effective crystal radii, $R_C$, of $M^{2+}$ cations. (Right) Correlation of intralayer Ag'-S-Ag'' angle (α) with $J_{2D}$ exchange constant in $MAg(SO_4)_2$, the dashed line shows a logarithmic fit of the form $J_{2D} = a \cdot \ln(\alpha - \alpha_0) + b$ with $\alpha_0 = 132.6°$.

**Table 4.** Summary of the one known and nine hypothetical quasi-ternary systems based on $MAg(SO_4)_2$ structure type of $P2_1/n$ space group.

| Cation | $R_C$ [Å] | ΔE [kJ/mol] | α [°] | $J_{2D}$ [meV] |
|---|---|---|---|---|
| $Zn^{2+}$ | 0.88 | –11.3 | 134.6 | –11.5 |
| $Cd^{2+}$ | 1.09 | –12.1 | 135.7 | –18.3 |
| $Hg^{2+}$ | 1.16 | 2.9 | 136.4 | –20.8 |
| $Mg^{2+}$ | 0.86 | –19.3 | 134.9 | –12.7 |
| $Ca^{2+}$ | 1.14 | 11.1 | 138.1 | –30.3 |
| $Sr^{2+}$ | 1.32 | 18.3 | 149.6 | –42.3 |
| $Ba^{2+}$ | 1.49 | 28.8 | 166.8 | –48.9 |
| $Cu^{2+}$ | 0.87 | –7.7 | 134.0 | –11.1 |
| $Ni^{2+}$ | 0.83 | –7.4 | 134.3 | –11.0 [a] |
| $Co^{2+}$ | 0.89 | –8.6 | 134.5 | –12.4 [a] |

[a] Indicated values of $J_{2D}$ are predicted from the logarithmic function fit.

## Conclusions

Copper(II) silver (II) sulfate was synthesized starting from binary fluorides and sulfuric acid. The title compound has the same crystal structure as the already known mineral $MgCu(SO_4)_2$ while both correspond to a monoclinic distortion of the $CuSO_4$ structure. The new salt features $[Ag(SO_4)_2]^{2-}$ layers, while the remaining cationic sites may be occupied either completely by $Cu^{2+}$ cations or partially by $Cu^{2+}$ and $Ag^{2+}$, corresponding to the formula of $Cu_xAg_{1-x}(SO_4)_2$ for $x = 0.3 – 0.5$. $CuAg(SO_4)_2$ is a rare example of an $Ag^{2+}$ based 2D antiferromagnet having a negative Curie-Weiss temperature of –84.1 K. The observed weak ferromagnetism presumably comes from spin canting.
DFT calculations show that the strongest exchange interaction in $CuAg(SO_4)_2$ corresponds to 2D AFM coupling within $[Ag(SO_4)_2]^{2-}$ layers, with the $J_{2D}$ constant of –11.1 meV. The compound is a CT insulator with a strong mixing of *d* and *p* states in both valence and conduction bands. This covalence, induced by strong oxidizing properties of $Ag^{2+}$, leads to a partial free-radical character of $SO_4^{2-}$ dianions which facilitate SE coupling between $Ag^{2+}$ sites. The magnetism of $CuAg(SO_4)_2$ is worth to be investigated in more detail using neutron scattering and muon spin spectroscopy.
Finally, relying on quantum mechanical calculations, we predict that $Cu^{2+}$ sites in $CuAg(SO_4)_2$ can be substituted with other divalent cations while preserving the structural 2D character and leading to an even stronger antiferromagnetic exchange of up to –50 meV. This compositional flexibility is new to $Ag^{2+}$ systems, and it is nearly absent in the numerous fluorides. The formation of high-entropy multicomponent sulfate is thus naturally expected[56].

## Experimental and computational methods

Synthetic method: Handling of volatiles (anhydrous HF, $F_2$), nonvolatiles, and the type of reaction vessels made of fluorinated polymer have already been reported[57]. $CuF_2$ (Aldrich, 98%) was treated with elemental $F_2$ at 220 °C for several hours before use. $AgF_2$ was freshly prepared by fluorination of $AgNO_3$ by elemental fluorine (Solvay Fluor and Derivate GmbH, 99.98%) in anhydrous HF (Linde AG, Pullach, Germany, 99.995%) at RT. Solid reagents $AgF_2$ and $CuF_2$ (**Table S2**) were loaded into a reaction vessel in a dry box. The reaction vessel was connected to a vacuum system and argon was pumped out. The mixtures of solid substrates were cooled to 77 K with liquid nitrogen and liquid anhydrous $H_2SO_4$ (2 ml) was slowly poured onto the mixture (**Figure S3**). The reaction mixture was warmed to RT and placed on an orbital shaker for 1-3 days. Finally, the resulting very viscous suspension was rinsed several times (5-9) with a large amount of HF (22-24 ml) to wash away the excess $H_2SO_4$. In the end, the last traces of HF were pumped off on a vacuum system.

The powder XRD measurements were conducted at RT using PANalytical X'Pert Pro diffractometer, with a linear PIXcel Medipix2 detector and no monochromator (a parallel beam of $CoK_{\alpha1}$ and $CoK_{\alpha2}$ with an intensity ratio of 2:1). Since samples easily decompose in contact with air, each sample was enclosed inside a sealed quartz capillary during XRD measurement. The structure was solved starting from the previously published model[19]. The final Rietveld refinement[27] for $CuAg(SO_4)_2$ was done using GSAS-II software[28]. Details of this structure refinement include: (1) 32-term Chebyshev polynomial background; (3) refined isotropic ADP $U_{iso}$ equal for atoms O1, O2, O3, O4; (4) refined anisotropic ADP $U_{aniso}$ equal for sites Ag1 and Cu1; (5) restrained distances targeted at 1.480 Å (esd. 0.010) for all S-O bonds. The absorption correction of the Debye-Scherer (cylinder) type was also refined with the least-square method, which resulted in $\mu_r = 4.1$ $mm^{-1}$. Constrained Rietveld refinements of intermediate $Cu_xAg_{1-x}SO_4$ phases were conducted using Jana2006 software[58].

Magnetic response measurements were conducted using a Quantum Design MPMS SQUID VSM magnetometer. The magnetization was measured between 2 and 300 K in different applied magnetic field regimes. The ZFC susceptibility was measured in a 1 kOe magnetic field in temperatures from 2 K to 300 K, after cooling down the sample in a zero magnetic field starting from 300 K. The FC susceptibility was measured in 1 kOe field magnetic in temperatures from 2 K to 300 K, after cooling down the sample in 1 kOe magnetic field. For these measurements, powder samples with a mass of 15-30 mg were loaded and closed inside PTFE tubes under a dry argon atmosphere and transported into the measurement chamber. The raw data were corrected for the contribution of an empty sample holder and the diamagnetic response of the constituent atoms using Pascal's constants[59].

Spin-polarized periodic DFT calculations were conducted with the projected augmented wave method[60] as implemented in Vienna Ab initio Simulation Package 5.4.4 (VASP)[61]. For the DFT+U formalism, incorporating the strong on-site interactions characteristic of open-shell transition metals, we used generalized gradient approximation (GGA) exchange-correlation functional PBEsol[62] together with the approach by Liechtenstein et al.[63]. The on-site interactions were included using the electrostatic Hubbard term $U$ set for Ag (5 eV, $l$ = 2), Cu (9 eV, $l$ = 2), S (2 eV, $l$ = 1), and O (4 eV, $l$ = 1), and the exchange term $J$ set equally for all these atoms (1 eV). For the spin-polarized DFT+U calculations, the k-spacing was set to 0.15 $Å^{-1}$. A similar computational setup can be found in other studies of silver(II) salts[18,19,26,31,34]. For the selected cases, we also used DFT methods without the above semi-empirical parameters. The first reference method used is spin-polarized hybrid DFT using HSE06 functional[64], with the k-spacing set to 0.30 $Å^{-1}$. The second reference method used is spin-polarized meta-GGA functional SCAN[65], with the k-spacing set to 0.20 $Å^{-1}$. For all the methods, the plane-wave energy cut-off was set to 520 eV. The electronic band gaps were calculated using the tetrahedron method with Blöchl corrections. VESTA[66] software was used for crystal structure visualization.

## Supporting Information

The data supporting the findings of this study are available in the electronic supplementary material of this article.

## Acknowledgments

This research was supported by the Polish National Science Center (NCN) within WG's Maestro project (2017/26/A/ST5/00570). The research was carried out with the use of CePT infrastructure financed by the European Union – the European Regional Development Fund within the Operational Programme "Innovative economy" for 2007–2013 (POIG.02.02.00-14-024/08-00). ZM acknowledges the financial support of the Slovenian Research Agency (research core funding No. P1-0045; Inorganic Chemistry and Technology). Calculations have been carried out using resources provided by the Wroclaw Centre for Networking and Supercomputing (http://wcss.pl), and the Interdisciplinary Centre of Mathematical and Computational Modelling (grant SAPPHIRE GA83-34).

**Keywords:** silver • layered compounds • exchange interactions • magnetic properties • density functional calculations

# A unique two-dimensional silver(II) antiferromagnet $Cu[Ag(SO_4)_2]$ and perspectives for its further modifications

Mateusz Domański,*[a] Zoran Mazej,*[b] and Wojciech Grochala*[a]

## Contents

[a] Mateusz Domański, Prof. Dr. Wojciech Grochala
Center of New Technologies
University of Warsaw
Zwirki i Wigury 93, 02-089 Warsaw Poland
E-mail: m.domanski@cent.uw.edu.pl, w.grochala@cent.uw.edu.pl

[b] Dr. Zoran Mazej
Department of Inorganic Chemistry and Technology
Jožef Stefan Institute
Jamova cesta 39, SI-1000 Ljubljana Slovenia
E-mail: zoran.mazej@ijs.si

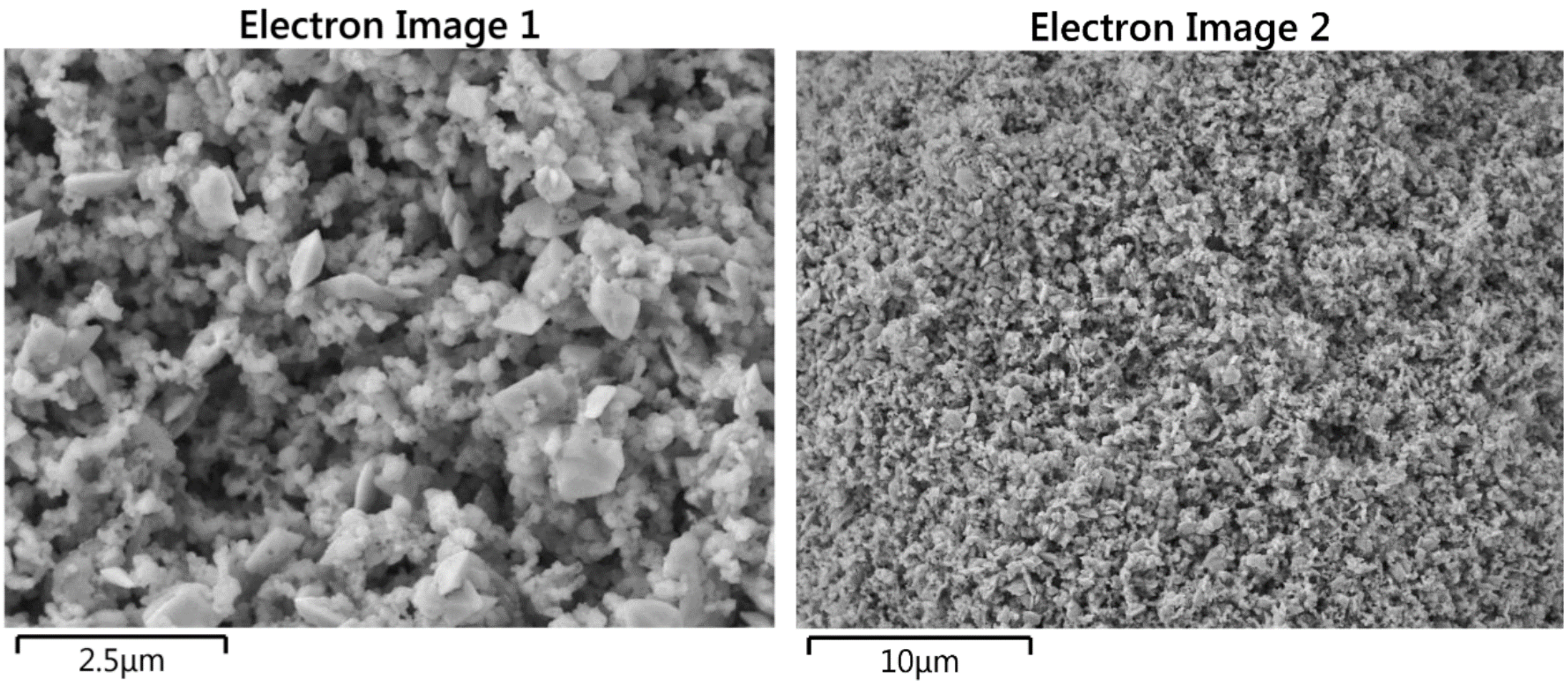

| Element | Line Type | Apparent Concentration | Wt% | Atomic % |
|---|---|---|---|---|
| O | K series | 75.74 | 25.25 | 59.06 |
| S | K series | 73.58 | 16.71 | 19.50 |
| Cu | L series | 15.78 | 5.39 | 3.17 |
| Ag | L series | 174.59 | 52.65 | 18.26 |
| Total: | | | 100.00 | 100.00 |

mol Cu/(Ag+Cu) = 14.8%

**Figure S1.** Scanning electron microscope images of sample **2** described in the previous publication[1], together with energy-dispersive X-ray spectroscopy (EDXS) elemental analysis. The powder XRD refinement resulted in about 57% molar concentration of $P2_1/n$ phase, with a concentration of copper of about 30% according to the volume-concentration relation as shown in **Figure S7**, which is in good agreement with the result above.

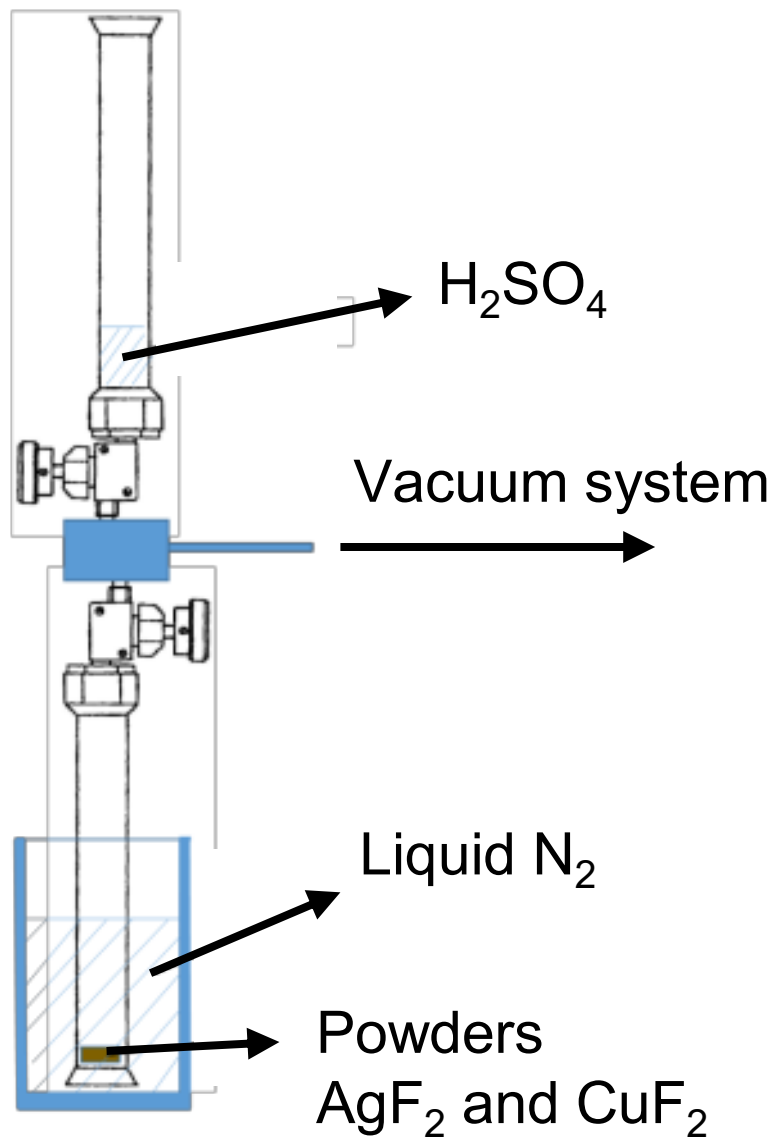

**Figure S2.** Experimental setup for the reaction between $AgF_2$, $CuF_2$, and $H_2SO_4$, similar to a setup presented in our previous work[1]. The sulfuric acid was added dropwise to the solid substrates to prevent a vigorous reaction.

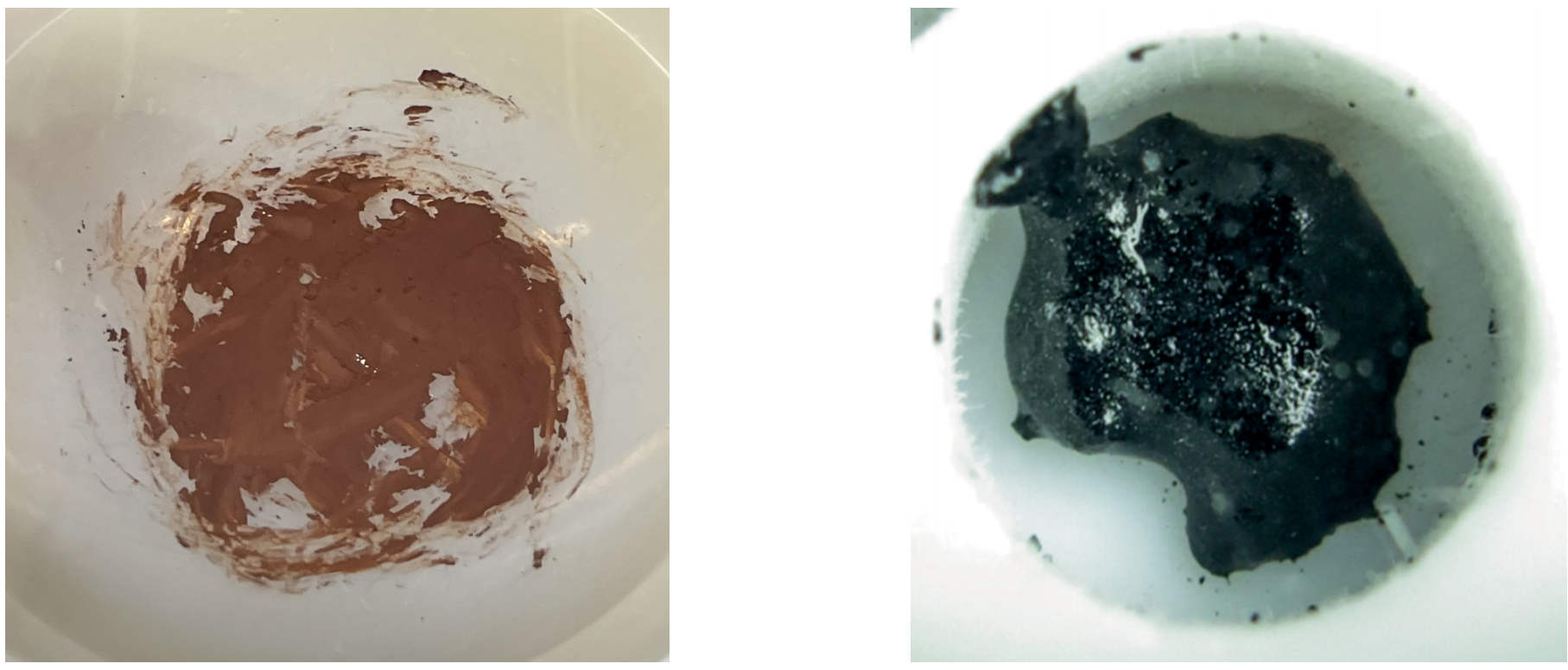

**Figure S3.** Ground sample of pure $CuAg(SO_4)_2$ in an agate mortar (left), showing its brown color, which significantly differs from black $AgSO_4$ (right, photo from the work by Połczyński et al.[2]) and white anhydrous $CuSO_4$.

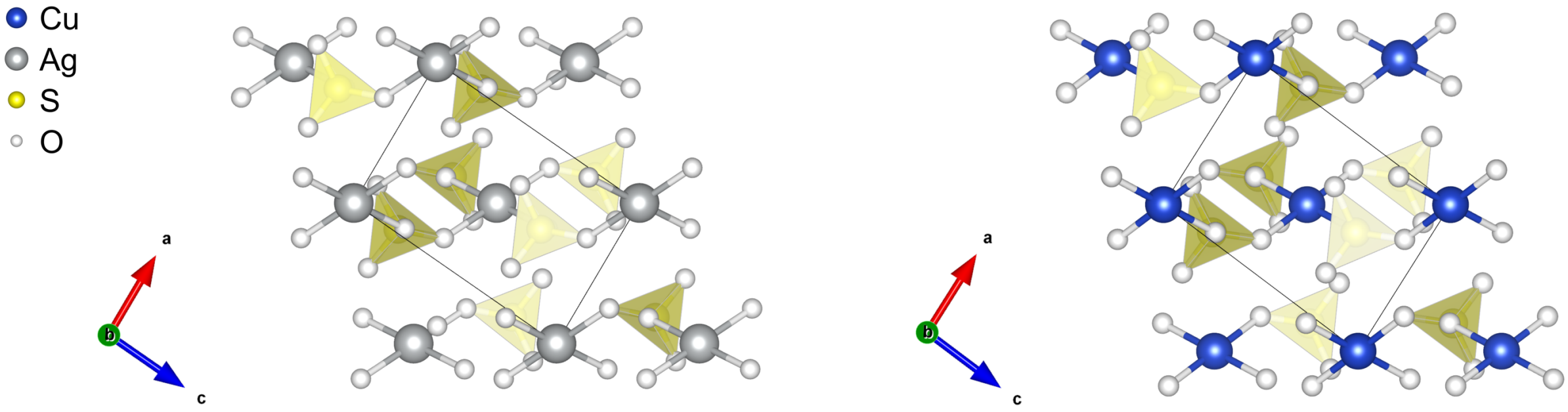

**Figure S4.** A view parallel to the $M(SO_4)_2^{2-}$ layers present in $CuAg(SO_4)_2$ and $MgCu(SO_4)_2$ crystal structure. The layers can be found parallel to (–110) planes of the unit cells.

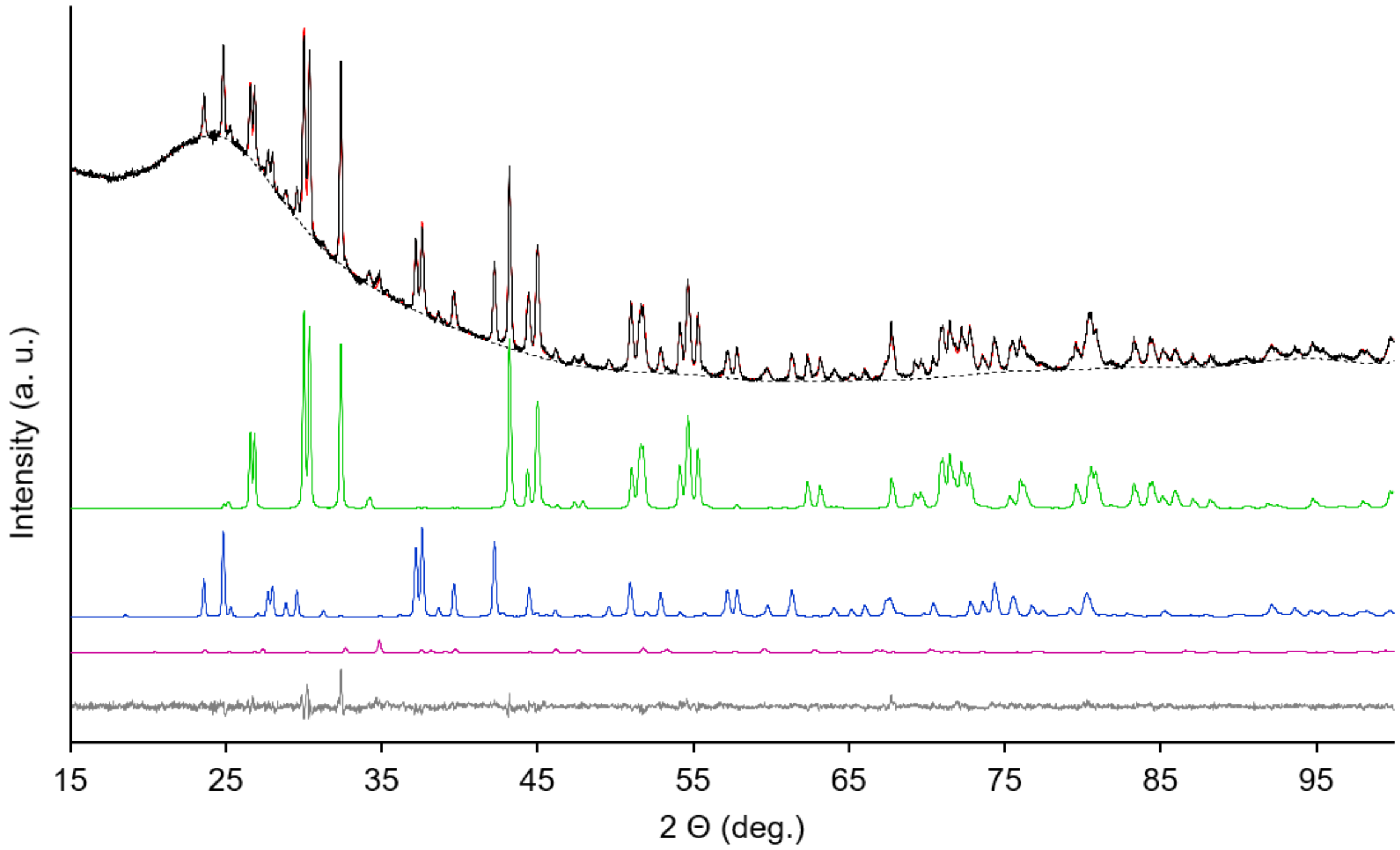

**Figure S5.** PXRD pattern of the sample with a starting ratio 85(Ag):15(Cu) (black line) and the calculated pattern (red), together with the polynomial background (dashed black line). Profiles of $AgSO_4$ (green), $Cu_{0.41}Ag_{0.59}SO_4$ (blue), and $AgSO_4H$ (magenta) calculated using Rietveld refinement are shown below. The differential profile is at the bottom (gray line). The Cu concentration in the $P2_1/n$ phase was determined from the Rietveld refinement by partial substitution of the Cu1 site with Ag, and refining the substitution coefficient together with the pattern, which resulted in 41.2(8)%. Fit parameters are wRp = 1.49% (pattern), R(obs) = 2.52% for $AgSO_4$, R(obs) = 2.69% for $Cu_{0.41}Ag_{0.59}SO_4$ and R(obs) = 4.05 for $AgSO_4H$. Phase analysis yields 58.5(3)% $AgSO_4$, 39.0(9)% $Cu_{0.41}Ag_{0.59}SO_4$ and 2.49(8)% $AgSO_4H$ in mass, which is molar composition of 56.4(3)% $AgSO_4$, and 41.3(9)% and 2.4(1)% $AgSO_4H$, which results 16.9(4)% mol Cu in the sample. This result inconsiderably deviates from the composition of the starting mixture of a molar ratio of 85(Ag):15(Cu).

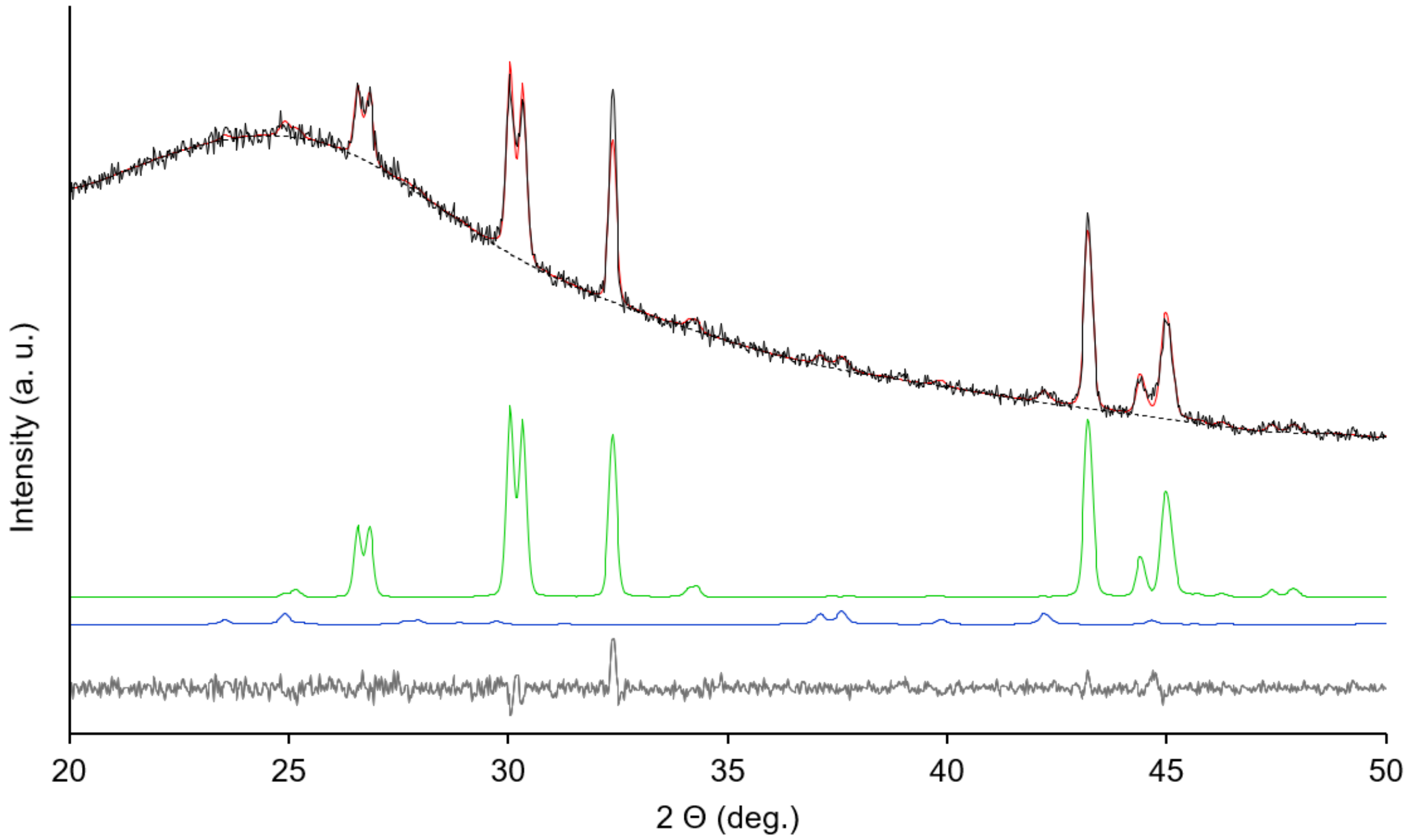

**Figure S6.** PXRD pattern of sample 95(Ag):5(Cu) (black solid line) and the calculated pattern (red), together with the polynomial background (dashed line). Profiles of $AgSO_4$ (green) and $Cu_{0.41}Ag_{0.59}SO_4$ (blue) calculated using Rietveld refinement are shown below. The Cu concentration in the $P2_1/n$ phase was roughly estimated from the unit cell volume. The differential profile is at the bottom (black). Fit parameters are wRp = 2.24% (pattern), R(obs) = 4.28% for $AgSO_4$, R(obs) = 7.65% for $Cu_{0.4}Ag_{0.6}SO_4$. Phase analysis yields 88.3(10)% $AgSO_4$ and 11.7(10)% $Cu_{0.41}Ag_{0.59}SO_4$, which is molar composition of 87.3(10)% $AgSO_4$ and 12.7(10)% $Cu_{0.41}Ag_{0.59}SO_4$ in mass, which yield 5.1(10)% mol Cu in the sample. This result is consistent with the starting mixture composition with a molar ratio of 95(Ag):5(Cu).

**Table S1.** Comparison of different levels of theory in the case of $CuSO_4$:$AgSO_4$ system in terms of volume prediction. The volume of mixing is defined as $V_{mix} = V_{P21/n} - (V_{C2/c} + V_{Pnma})/2$ per formula unit (FU).

| Method | Volume [Å$^3$] per FU | | | $V_{exp}$ / $V_{theor}$ | | | $V_{mix}$ [Å$^3$] | $V_{prod}$ / $V_{substr}$ |
|---|---|---|---|---|---|---|---|---|
| | $AgSO_4$ | $Cu_{0.5}Ag_{0.5}SO_4$ | $CuSO_4$ | $AgSO_4$ | $Cu_{0.5}Ag_{0.5}SO_4$ | $CuSO_4$ | | |
| Exp. (XRD) | 75.88 | 73.67 | 68.17 | - | - | - | +1.65 | 102.3% |
| GGA+U | 78.77 | 73.19 | 65.60 | 96.3% | 100.7% | 103.9% | +1.00 | 101.4% |
| HSE06 | 78.14 | 73.48 | 66.25 | 97.1% | 100.3% | 102.9% | +1.29 | 101.8% |
| SCAN | 77.13 | 73.79 | 67.06 | 98.4% | 99.8% | 101.7% | +1.70 | 102.4% |

**Table S2.** Summary of the samples, obtained in the reaction between $CuF_2$, $AgF_2$, and $H_2SO_4$ in the attempts to prepare $Cu_xAg_{1-x}SO_4$ salts. The volume is provided for 4 cations in a unit cell to easily compare with quasi-binary systems. The $x$ variable denotes Cu concentration in $Cu_xAg_{1-x}SO_4$ $P2_1/n$ phase. The $x$ was obtained from constrained Rietveld powder refinement together with unit cell parameters. We note that $x$ strongly correlates with the linear absorption parameter, thus, the results shown are provided for the value which resulted in the lowest wRp fit parameter.

| Sample | Quantities used n($CuF_2$) : n($AgF_2$) [mmol] | Present phases | $x$ | Volume per Z = 4 [Å$^3$] |
|---|---|---|---|---|
| Impure substrate | - | $Cu_xAg_{1-x}SO_4$, $AgSO_4$, $AgSO_3F$, low level of impurities | 31.3(9)% | 299.59 |
| Impure substrate | - | $Cu_xAg_{1-x}SO_4$, $AgSO_3F$, low level of impurities | 38.9(11)% | 299.44 |
| Impure substrate | - | $Cu_xAg_{1-x}SO_4$, high level of impurities | 35.8(7)% | 297.96 |
| Impure substrate | - | $Cu_xAg_{1-x}SO_4$, $AgSO_4$, high level of impurities | 30% [b] | 300.13 |
| Pure fluorides, Cu:Ag 50:50 | 1.10 : 1.10 | $Cu_xAg_{1-x}SO_4$ | 50% [a] | 294.67 |
| Pure fluorides, Cu:Ag 15:85 | 0.49 : 2.75 | $Cu_xAg_{1-x}SO_4$, $AgSO_4$, $AgSO_4H$ | 41.2(8)% | 298.36 |
| Pure fluorides, Cu:Ag 5:95 | 0.10 : 1.88 | $Cu_xAg_{1-x}SO_4$, $AgSO_4$ | 41% [b] | 297.68 |
| $AgSO_4$ (*C2/c*, CIF[3]) | - | - | 0% | 303.51 |
| $CuSO_4$ (*Pnma*, CIF[4]) | - | - | 100% | 272.66 |

[a] the Cu concentration in this sample was assumed according to the Ag:Cu ratio in the substrates used.
[b] the Cu concentration in this sample was assumed based on the volume-to-concentration correlation, **Figure S7**.

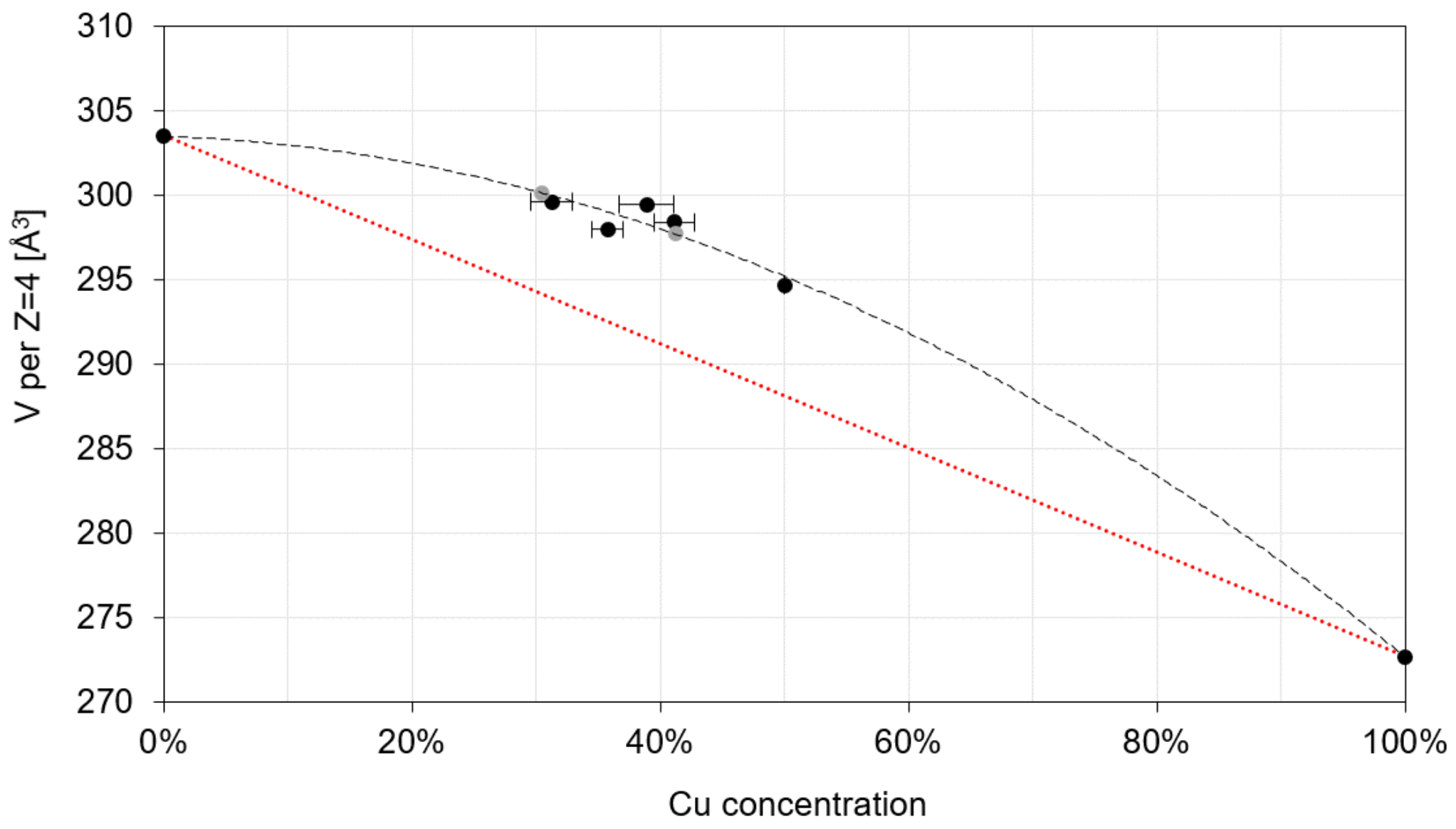

**Figure S7.** Plotted data from **Table S2** shows deviation from Vegard's law for solid solutions. The uncertainties of Cu concentrations come from the least-squares Rietveld refinement pattern and are doubled. Grey circles show the concentrations which could not be estimated experimentally and are fitted to the correlation curve of 2nd degree polynomial.

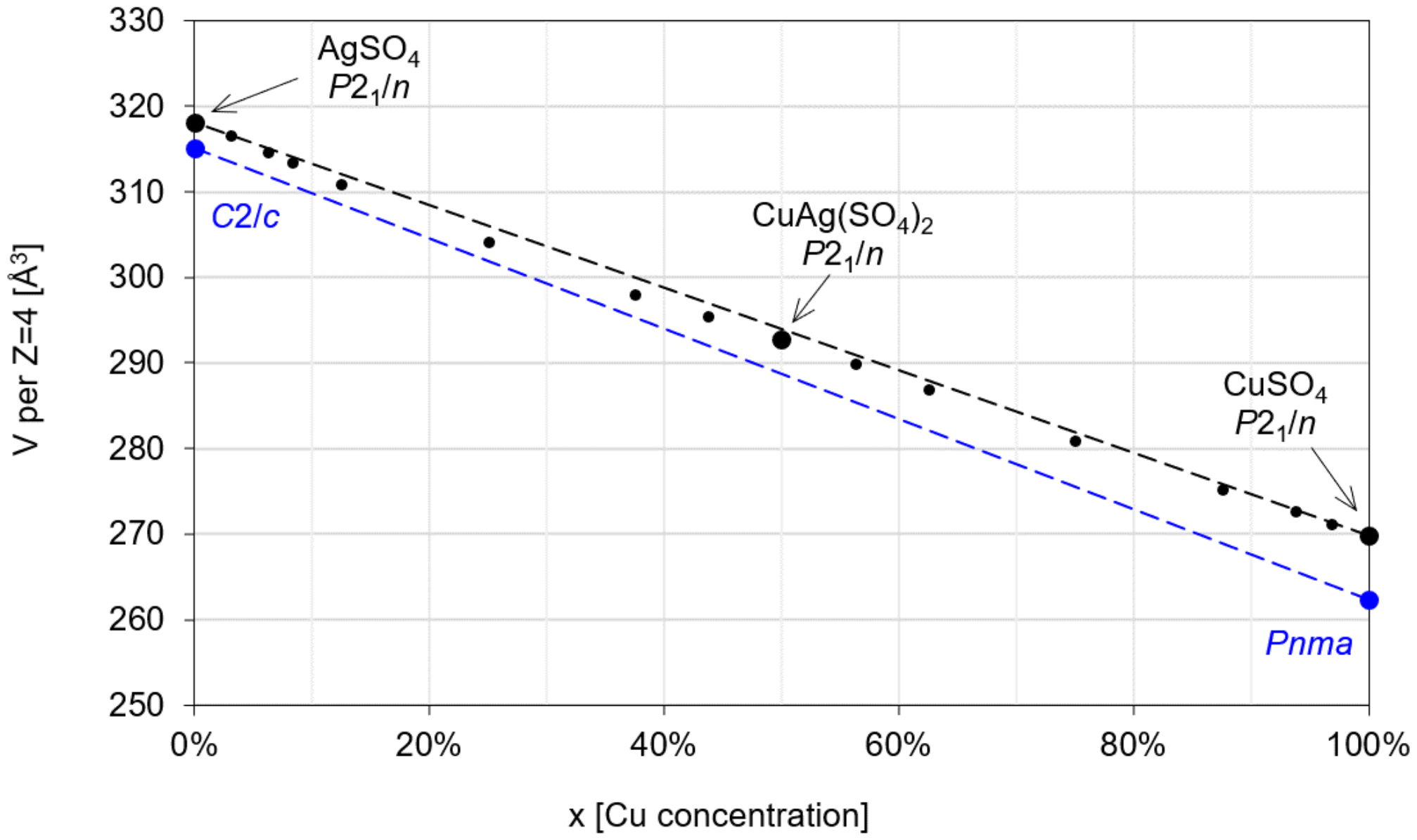

**Figure S8.** Relation between the volume of quasi-binary substrates and intermediate stoichiometries $Cu_xAg_{1-x}SO_4$ calculated with the GGA+U method. Dashed lines indicate intermediate volumes according to Vegard's law between experimentally known polymorphs (blue line) and hypothetical *P*2₁/*n* polymorphs (black line) of the binary sulfates.

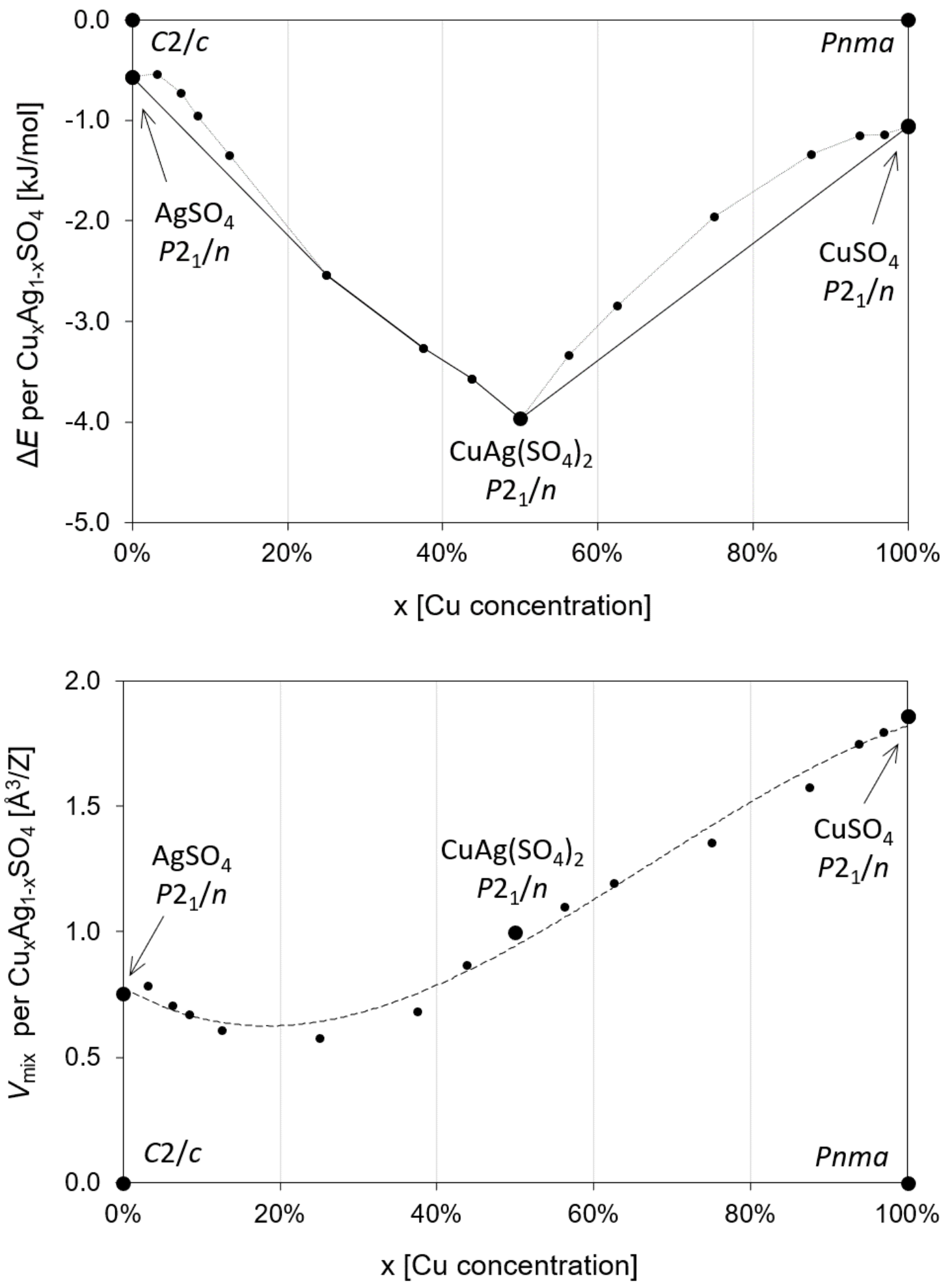

**Figure S9.** Relation between quasi-binary substrates and intermediate stoichiometries $Cu_xAg_{1-x}SO_4$ calculated with GGA+U method. **Top:** the electronic energy of the reaction $\mathrm{x}\,\mathrm{CuSO_4}(Pnma) + (1-\mathrm{x})\,\mathrm{AgSO_4}(C2/c) \rightarrow \mathrm{Cu}_x\mathrm{Ag}_{1-\mathrm{x}}\mathrm{SO_4}(P2_1/n)$, *i.e.* in relation to experimentally known forms of quasi-binary substrates. The full lines indicate stable stoichiometries (convex hull), while dotted lines only catch the eye. It should be noted that the GGA+U method tends to overestimate $P2_1/n$ phase stability in comparison to higher-level DFT methods. **Bottom:** the excess volume of mixing for various Cu substitution levels in $P2_1/n$ crystal structure, the dashed line shows a 3rd degree polynomial correlation line.

**Table S3.** Comparison of different levels of theory in the case of $CuSO_4$:$AgSO_4$ system in terms of electronic energy. The energy of phase transition per mole, $\Delta E_{PT}$ (in kJ/mol), is calculated as a difference between the $P2_1/n$ space group and the experimentally known one. The energy of formation of the ternary compound, $\Delta E_f$ (in kJ/mol), is the energy of reaction $0.5\,\mathrm{CuSO_4}(Pnma) + 0.5\,\mathrm{AgSO_4}(C2/c) \rightarrow \mathrm{Cu_{0.5}Ag_{0.5}SO_4}(P2_1/n)$, i.e. calculated per mole of sulfate groups.

| Method | $AgSO_4$ | | $CuSO_4$ | | $Cu_{0.5}Ag_{0.5}SO_4$ | |
|---|---|---|---|---|---|---|
| | *C*2/*c* | *P*2$_1$/*n* | *Pnma* | *P*2$_1$/*n* | *P*2$_1$/*n* | |
| | $\Delta E_{PT}$ | $\Delta E_{PT}$ | $\Delta E_{PT}$ | $\Delta E_{PT}$ | $\Delta E_f$ | $E_{gap}$ |
| GGA+U | 0 | –0.26 | 0 | –1.06 | –3.97 | 1.364 |
| HSE06 | 0 | 0.56 | 0 | –2.21 | –3.69 | 2.231 |
| SCAN | 0 | 2.78 | 0 | 0.16 | –1.21 | 0.673 |

## Experimental methods for $CuAg(SO_4)_2$ vibrational spectroscopy

Infrared spectroscopy (IR) measurements of the $CuAg(SO_4)_2$ and $CuSO_4$ samples were carried out using a Bruker Vertex 80V spectrometer (Bruker, Billerica, MA, USA). For the FIR range measurements, pure fine powdered samples were placed on HDPE windows; in the case of the MIR range measurement anhydrous KBr stored in a dry box was used as a dispersing material. MIR and FIR spectra were merged at about 500 $cm^{-1}$ for $CuSO_4$ and 550 $cm^{-1}$ for $CuAg(SO_4)_2$, at the absance of strong bands. Raman spectra were collected with Horiba Jobin Yvon Raman spectrometer with 532 nm Nd:YAG laser exciting beam with 5.0 mW power for samples in sealed quartz capillaries. Similar to $AgF_2$ and $AgSO_4$, $CuAg(SO_4)_2$ is relatively sensitive to visible light, and it decomposes in the 532 nm laser beam, which is noticeable with the increasing intensity of the 970 $cm^{-1}$ band coming from $Ag(I)_2SO_4$ photochemical product when laser power is increased. The $Ag_2SO_4$ signal becomes visible already if radiated longer than 30 min with 5.0 mW power, with white spots appearing on the powder surface. For this reason, the capillary with $CuAg(SO_4)_2$ was slowly moved in the beam (at about 1 mm/h).

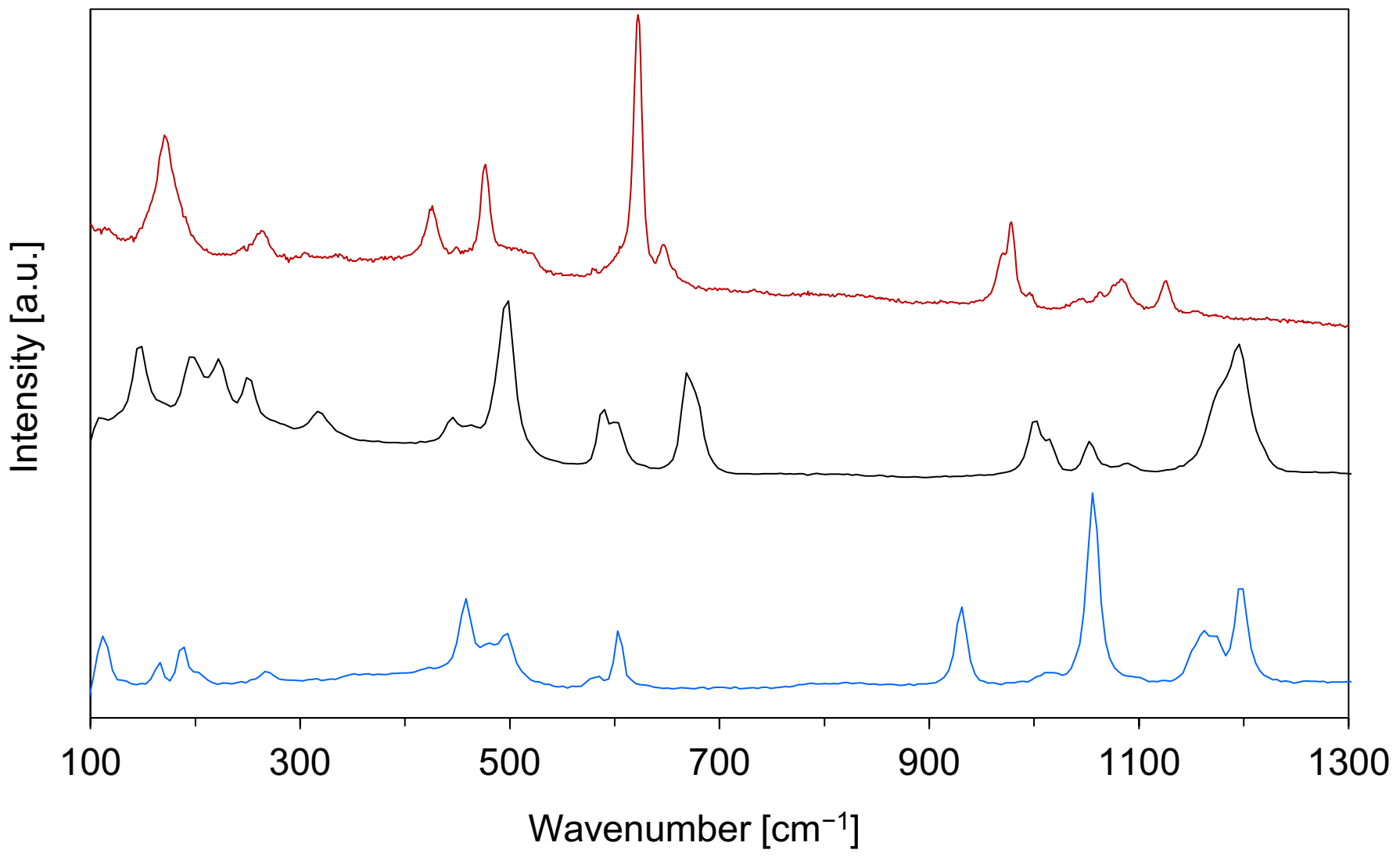

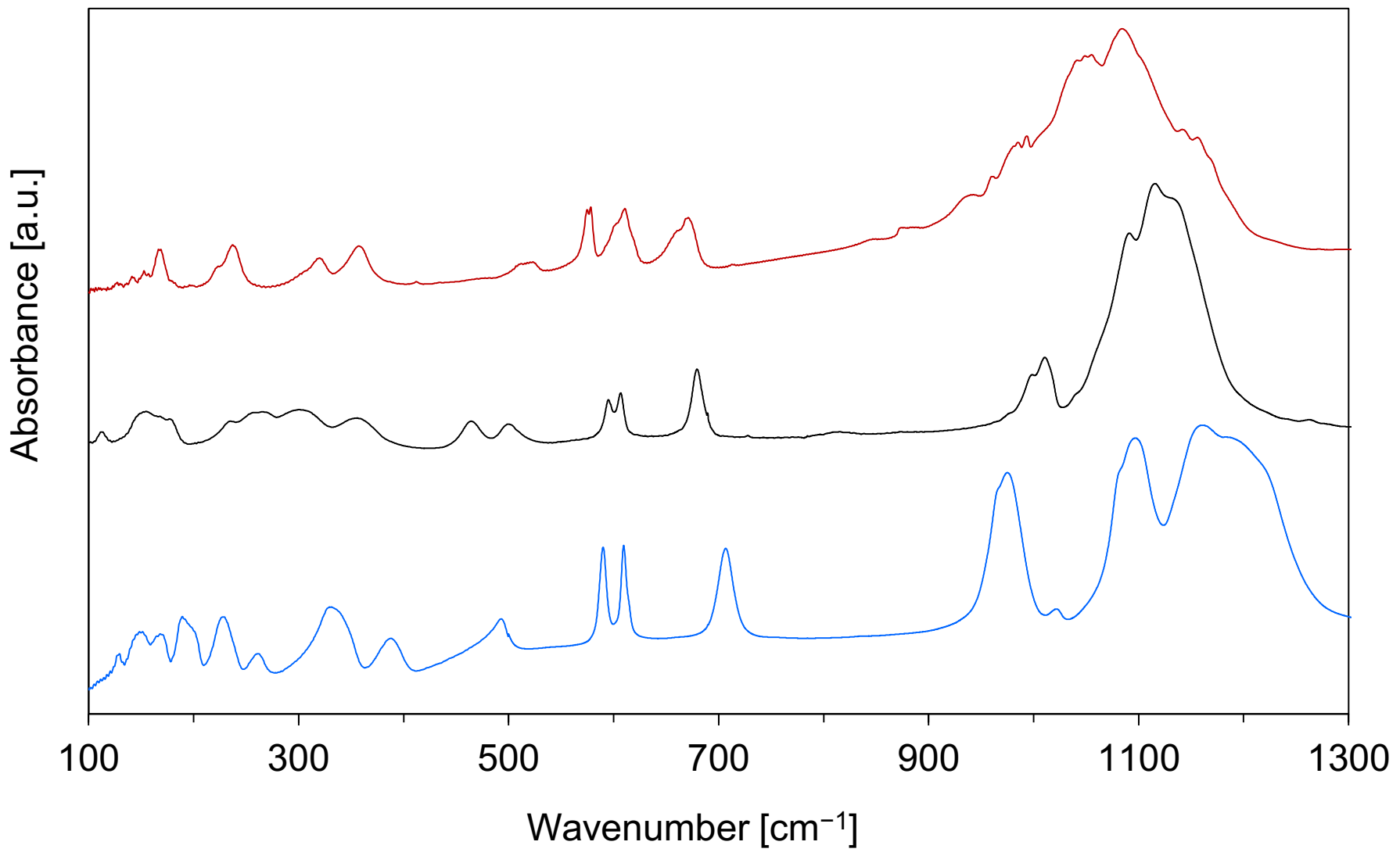

**Figure S10.** IR absorbance and Raman scattering spectra of $AgSO_4$ (red), $CuAg(SO_4)_2$ (black), and $CuSO_4$ (blue). (Top) IR spectra are combined from MIR and FIR parts. (Bottom) Raman spectra were collected using the 532 nm line for $CuAg(SO_4)_2$ and $CuSO_4$ and the 632 nm line for $AgSO_4$. The spectra of $AgSO_4$ come from the work by Połczyński et al.[2]

**Table S4.** Bands observed for $CuSO_4$, $CuAg(SO_4)_2$, and $AgSO_4$ with IR absorption and Raman scattering vibrational spectroscopy. Wavenumbers are provided in $cm^{-1}$ together with their relative apparent strength and estimated vibrational assignment.

| $CuSO_4$ | | | $CuAg(SO_4)_2$ | | | $AgSO_4$ | | |
|---|---|---|---|---|---|---|---|---|
| Raman | IR | | Raman | IR | | Raman | IR | |
| $\tilde{\nu}$ | $\tilde{\nu}$ | Mode | $\tilde{\nu}$ | $\tilde{\nu}$ | Mode | $\tilde{\nu}$ | $\tilde{\nu}$ | Mode |
| 112 mw | | r $[SO_4]^{2-}$ | 108 vw | 112 vw | r $[SO_4]^{2-}$ | 171 vs | 167 mw | r $[SO_4]^{2-}$ |
| | 129 vw | r $[SO_4]^{2-}$ | 149 ms | 155 w, b | r $[SO_4]^{2-}$ | | 224 sh | $\nu$ [Ag-O] |
| | 148 w | r $[SO_4]^{2-}$ | | 177 w, b | r $[SO_4]^{2-}$ | | 237 m | $\nu$ [Ag-O] |
| 167 w | 168 w | r $[SO_4]^{2-}$ | 195 m | | $\nu$ [M-O] | 264 m | | $\nu$ [Ag-O] |
| 189 mw | 189 mw | $\nu$ [Cu-O] | 222 m | | $\nu$ [M-O] | | 320 mw | $\nu$ [Ag-O] |
| | 201 sh | $\nu$ [Cu-O] | | 235 w, b | $\nu$ [M-O] | | 358 m | $\nu$ [Ag-O] |
| | 229 m | $\nu$ [Cu-O] | 249 m | | $\nu$ [M-O] | 426 s | | $\delta$ $[SO_4]^{2-}$ |
| 266 w | 261 w | $\nu$ [Cu-O] | | 258 w, b | $\nu$ [M-O] | 477 s | | $\delta$ $[SO_4]^{2-}$ |
| | 331 s | $\nu$ [Cu-O] | | | $\nu$ [M-O] | | 515 w | $\delta$ $[SO_4]^{2-}$ |
| | 388 m | $\nu$ [Cu-O] | | 266 w, b | $\nu$ [M-O] | 522 w | 523 w | $\delta$ $[SO_4]^{2-}$ |
| 458 ms | | $\delta$ $[SO_4]^{2-}$ | | 300 w, b | $\nu$ [M-O] | | 575 ms | $\delta$ $[SO_4]^{2-}$ |
| 480 w | | $\delta$ $[SO_4]^{2-}$ | 317 m | | $\nu$ [M-O] | 580 vw | 578 ms | $\delta$ $[SO_4]^{2-}$ |
| 498 mw | 493 mw | $\delta$ $[SO_4]^{2-}$ | | 356 mw, b | $\nu$ [M-O] | | 601 sh | $\delta$ $[SO_4]^{2-}$ |
| 585 w | 590 s | $\delta$ $[SO_4]^{2-}$ | 446 mw | | $\delta$ $[SO_4]^{2-}$ | | 611 s | $\delta$ $[SO_4]^{2-}$ |
| 603 ms | | $\delta$ $[SO_4]^{2-}$ | 463 w | 464 mw | $\delta$ $[SO_4]^{2-}$ | 622 vs | | $\delta$ $[SO_4]^{2-}$ |
| | 610 s | $\delta$ $[SO_4]^{2-}$ | 499 vs | 499 mw | $\delta$ $[SO_4]^{2-}$ | 647 mw | | $\delta$ $[SO_4]^{2-}$ |
| | 707 s | $\delta$ $[SO_4]^{2-}$ | 590 ms | 595 m | $\delta$ $[SO_4]^{2-}$ | | 662 sh | $\delta$ $[SO_4]^{2-}$ |
| 931 s | | $\nu$ [S–O] | 603 m | 607 ms | $\delta$ $[SO_4]^{2-}$ | | 671 s | $\delta$ $[SO_4]^{2-}$ |
| | 967 sh | $\nu$ [S–O] | 668 vs | | $\delta$ $[SO_4]^{2-}$ | 971 sh | | $\nu$ [S–O] |
| | 975 vs | $\nu$ [S–O] | | 679 s | $\delta$ $[SO_4]^{2-}$ | 978 s | | $\nu$ [S–O] |
| | 1022 w | $\nu$ [S–O] | 1003 s | 999 ms | $\nu$ [S–O] | | 990 s, b | $\nu$ [S–O] |
| 1056 vs | | $\nu$ [S–O] | 1015 m | 1011 s | $\nu$ [S–O] | 996 w | | $\nu$ [S–O] |
| | 1083 sh | $\nu$ [S–O] | 1052 m | | $\nu$ [S–O] | | 1040 vs, b | $\nu$ [S–O] |
| | 1097 vs | $\nu$ [S–O] | 1089 w | 1091 vs | $\nu$ [S–O] | 1063 vw | | $\nu$ [S–O] |
| 1163 ms | 1161 vs | $\nu$ [S–O] | | 1115 vs | $\nu$ [S–O] | 1084 m | 1085 vs, b | $\nu$ [S–O] |
| 1175 sh | | $\nu$ [S–O] | | 1134 vs | $\nu$ [S–O] | 1126 ms | | $\nu$ [S–O] |
| | 1187 vs | $\nu$ [S–O] | 1175 sh | | $\nu$ [S–O] | | 1150 s, b | $\nu$ [S–O] |
| 1195 s | | $\nu$ [S–O] | 1196 vs | | $\nu$ [S–O] | | | |
| | 1214 sh | $\nu$ [S–O] | | | | | | |

Intensities abbreviations code: v – very, s – strong, m –medium, w –weak, sh –shoulder, b – broad.
Assignment abbreviations code: $\nu$– bond stretching, $\delta$– bending, r – hindered rotation.

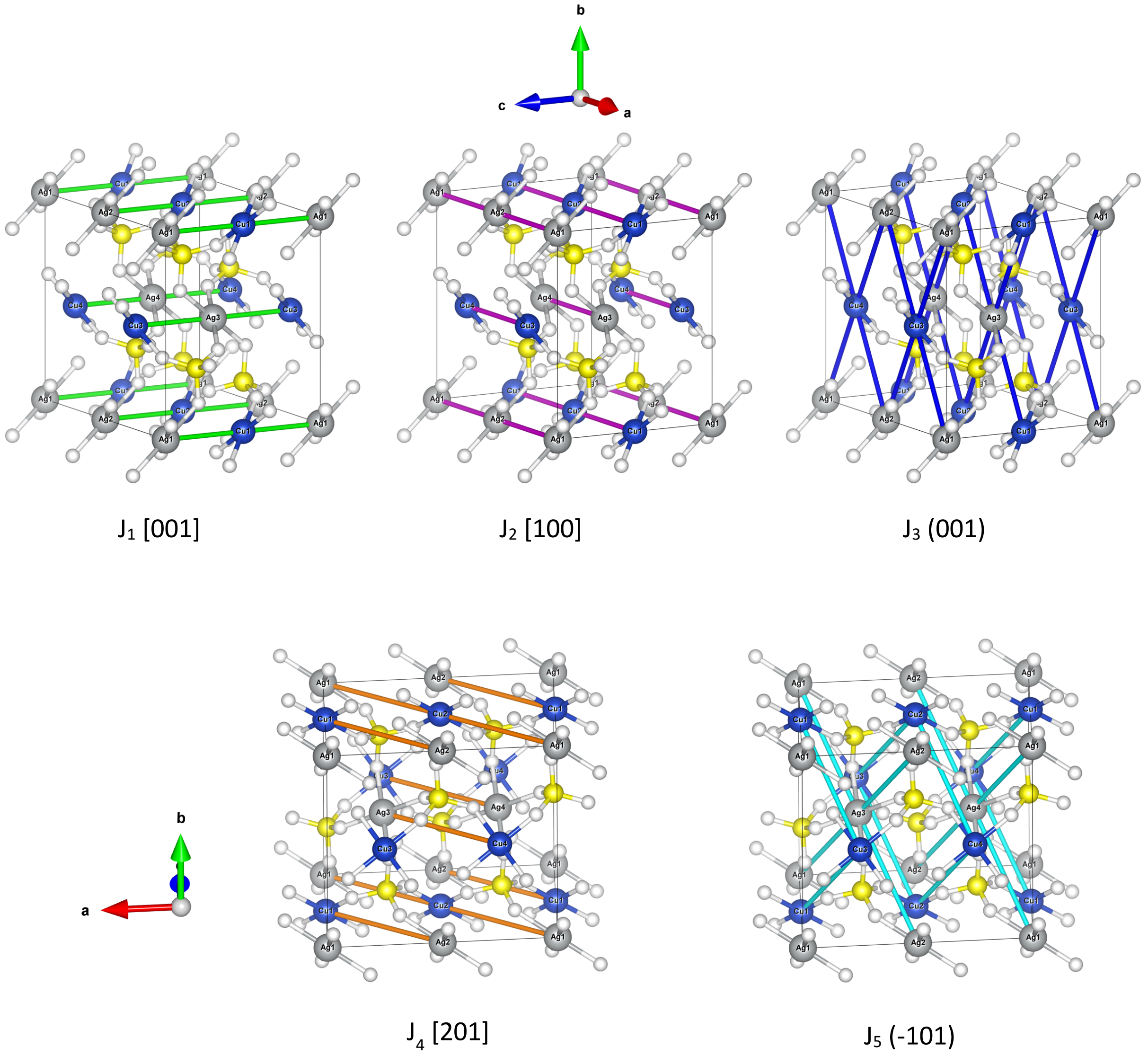

**Figure S11.** A general view of the possible magnetic interactions in $CuAg(SO_4)_2$ crystal structure. The cell edges and numeration of atoms are shown for 2x1x1 supercell, while the directions are shown with respect to the primitive unit cell. A detailed view of the exchange pathways is shown in **Figure S12**, while data and the results of calculations using **Equation S1** are shown in **Table S5** and **Table S6**.

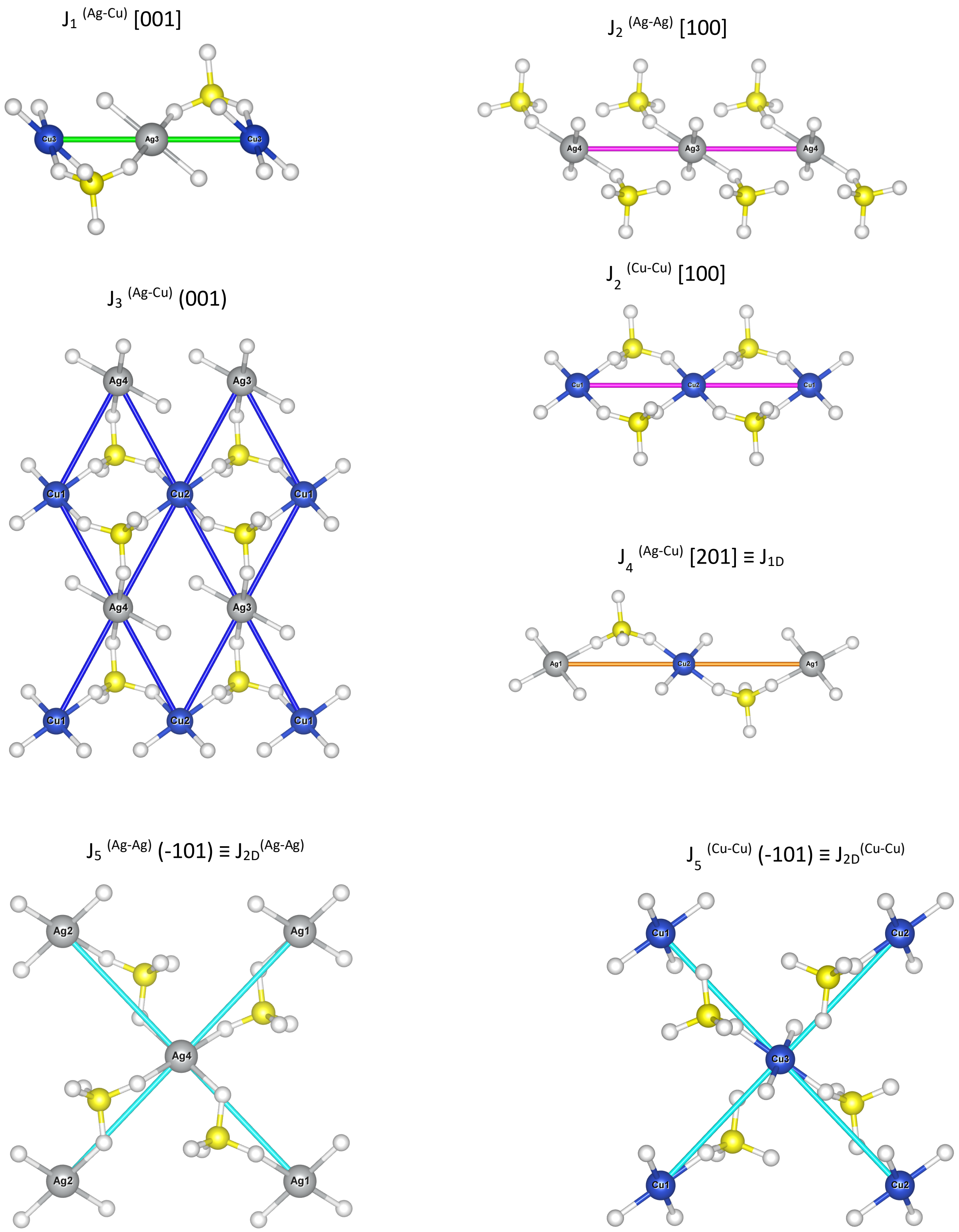

**Figure S12.** A detailed view of the possible magnetic interactions and their exchange pathways in $CuAg(SO_4)_2$ crystal structure. Only $SO_4^{2-}$ ligands that may involve in the interaction and are connected in-plane to $d(x^2–y^2)$ of $M^{2+}$ are shown. The cell edges and numeration of atoms are shown for 2x1x1 supercell, while the directions are shown with respect to the primitive cell.

**Equation S1.** The following Hamiltonian was used to calculate exchange interactions in the 2x1x1 supercell of $CuAg(SO_4)_2$ structure. The determinants method was used to solve systems of linear equations for each case, the calculated coupling constants are presented in **Table S5**.

$$
\begin{aligned}
H = & -2 \cdot J_1^{(Ag\text{-}Cu)} \cdot (S^{Ag1} \cdot S^{Cu1} + S^{Ag2} \cdot S^{Cu2} + S^{Ag3} \cdot S^{Cu3} + S^{Ag4} \cdot S^{Cu4}) \\
& -2 \cdot J_2^{(Ag\text{-}Ag)} \cdot (S^{Ag1} \cdot S^{Ag2} + S^{Ag3} \cdot S^{Ag4}) \\
& -2 \cdot J_2^{(Cu\text{-}Cu)} \cdot (S^{Cu1} \cdot S^{Cu2} + S^{Cu3} \cdot S^{Cu4}) \\
& -2 \cdot J_3^{(Ag\text{-}Cu)} \cdot ((S^{Ag1} + S^{Ag2}) \cdot (S^{Cu3} + S^{Cu4}) + (S^{Ag3} + S^{Ag4}) \cdot (S^{Cu1} + S^{Cu2})) \\
& -2 \cdot J_4^{(Ag\text{-}Cu)} \cdot (S^{Ag1} \cdot S^{Cu2} + S^{Ag2} \cdot S^{Cu1} + S^{Ag3} \cdot S^{Cu4} + S^{Ag4} \cdot S^{Ag3}) \\
& -2 \cdot J_{2D}^{(Ag\text{-}Ag)} \cdot ((S^{Ag1} + S^{Ag2}) \cdot (S^{Ag3} + S^{Ag4})) \\
& -2 \cdot J_{2D}^{(Cu\text{-}Cu)} \cdot ((S^{Cu1} + S^{Cu2}) \cdot (S^{Cu3} + S^{Cu4}))
\end{aligned}
$$

**Table S5.** The exchange constants energies for the supercell 2x1x1 of DFT optimized structures of $CuAg(SO_4)_2$ on different theory levels, including GGA+U, hybrid-DFT, and meta-GGA. Directions of interactions refer to the primitive cell of the $CuAg(SO_4)_2$. Energies calculated with DFT for selected spin states and spin eigenvalues are presented in **Table S6**.

| Interaction | Direction | E(GGA+U) [meV] | E(HSE06) [meV] | E(SCAN) [meV] |
|---|---|---|---|---|
| $J_1^{(Ag\text{-}Cu)}$ | [001] | 1.0 | 1.2 | 7.3 |
| $J_2^{(Ag\text{-}Ag)}$ | [100] | –0.2 | –0.2 | –1.0 |
| $J_2^{(Cu\text{-}Cu)}$ | [100] | 0.3 | 0.6 | 0.1 |
| $J_3^{(Ag\text{-}Cu)}$ | (001) | –0.1 | –0.2 | –0.3 |
| $J_4^{(Ag\text{-}Cu)}$ | [201] | –7.0 | –10.3 | –13.3 |
| $J_{2D}^{(Ag\text{-}Ag)}$ | (-101) | –9.1 | –11.1 | –14.4 |
| $J_{2D}^{(Cu\text{-}Cu)}$ | (-101) | 0.2 | 0.4 | 1.6 |

**Table S6.** Energy and $M^{2+}$ spin eigenvalues for the 2x1x1 supercell of DFT optimized structures of $CuAg(SO_4)_2$ on different theory levels. For both GGA+U and HSE06 the groundstate is AFM1 while for SCAN meta-GGA it is AFM5, both are shown in **Figure S13**.

a) GGA+U

| State | E(PBEsol+U) / eV | $S^{Ag1}$ | $S^{Ag2}$ | $S^{Ag3}$ | $S^{Ag4}$ | $S^{Cu1}$ | $S^{Cu2}$ | $S^{Cu3}$ | $S^{Cu4}$ |
|---|---|---|---|---|---|---|---|---|---|
| AFM1 (GS) | -196.987775 | -0.5 | -0.5 | 0.5 | 0.5 | 0.5 | 0.5 | -0.5 | -0.5 |
| AFM2 | -196.952784 | 0.5 | 0.5 | 0.5 | 0.5 | -0.5 | -0.5 | -0.5 | -0.5 |
| AFM3 | -196.964079 | 0.5 | 0.5 | -0.5 | -0.5 | 0.5 | 0.5 | -0.5 | -0.5 |
| AFM4 | -196.942089 | -0.5 | 0.5 | 0.5 | -0.5 | 0.5 | -0.5 | -0.5 | 0.5 |
| AFM5 | -196.973931 | -0.5 | 0.5 | 0.5 | -0.5 | -0.5 | 0.5 | 0.5 | -0.5 |
| FM4_1 | -196.976753 | -0.5 | -0.5 | 0.5 | 0.5 | 0.5 | 0.5 | 0.5 | 0.5 |
| FM4_3 | -196.939494 | 0.5 | 0.5 | 0.5 | 0.5 | 0.5 | -0.5 | 0.5 | -0.5 |
| FM8 | -196.928160 | 0.5 | 0.5 | 0.5 | 0.5 | 0.5 | 0.5 | 0.5 | 0.5 |

b) hybrid DFT

| State | E(HSE06) / eV | $S^{Ag1}$ | $S^{Ag2}$ | $S^{Ag3}$ | $S^{Ag4}$ | $S^{Cu1}$ | $S^{Cu2}$ | $S^{Cu3}$ | $S^{Cu4}$ |
|---|---|---|---|---|---|---|---|---|---|
| AFM1 (GS) | -359.493270 | -0.5 | -0.5 | 0.5 | 0.5 | 0.5 | 0.5 | -0.5 | -0.5 |
| AFM2 | -359.452283 | 0.5 | 0.5 | 0.5 | 0.5 | -0.5 | -0.5 | -0.5 | -0.5 |
| AFM3 | -359.458632 | 0.5 | 0.5 | -0.5 | -0.5 | 0.5 | 0.5 | -0.5 | -0.5 |
| AFM4 | -359.430689 | -0.5 | 0.5 | 0.5 | -0.5 | 0.5 | -0.5 | -0.5 | 0.5 |
| AFM5 | -359.476897 | -0.5 | 0.5 | 0.5 | -0.5 | -0.5 | 0.5 | 0.5 | -0.5 |
| FM4_1 | -359.477431 | -0.5 | -0.5 | 0.5 | 0.5 | 0.5 | 0.5 | 0.5 | 0.5 |
| FM4_3 | -359.431173 | -0.5 | -0.5 | -0.5 | -0.5 | -0.5 | 0.5 | -0.5 | 0.5 |
| FM8 | -359.413681 | 0.5 | 0.5 | 0.5 | 0.5 | 0.5 | 0.5 | 0.5 | 0.5 |

c) meta-GGA

| State | E(SCAN) / eV | $S^{Ag1}$ | $S^{Ag2}$ | $S^{Ag3}$ | $S^{Ag4}$ | $S^{Cu1}$ | $S^{Cu2}$ | $S^{Cu3}$ | $S^{Cu4}$ |
|---|---|---|---|---|---|---|---|---|---|
| AFM1 | -518.514673 | -0.5 | -0.5 | 0.5 | 0.5 | 0.5 | 0.5 | -0.5 | -0.5 |
| AFM2 | -518.465953 | 0.5 | 0.5 | 0.5 | 0.5 | -0.5 | -0.5 | -0.5 | -0.5 |
| AFM3 | -518.492875 | 0.5 | 0.5 | -0.5 | -0.5 | 0.5 | 0.5 | -0.5 | -0.5 |
| AFM4 | -518.438733 | -0.5 | 0.5 | 0.5 | -0.5 | 0.5 | -0.5 | -0.5 | 0.5 |
| AFM5 (GS) | -518.521140 | -0.5 | 0.5 | 0.5 | -0.5 | -0.5 | 0.5 | 0.5 | -0.5 |
| FM4_1 | -518.510301 | -0.5 | -0.5 | 0.5 | 0.5 | 0.5 | 0.5 | 0.5 | 0.5 |
| FM4_3 | -518.449292 | 0.5 | 0.5 | 0.5 | 0.5 | -0.5 | 0.5 | -0.5 | 0.5 |
| FM8 | -518.439732 | 0.5 | 0.5 | 0.5 | 0.5 | 0.5 | 0.5 | 0.5 | 0.5 |

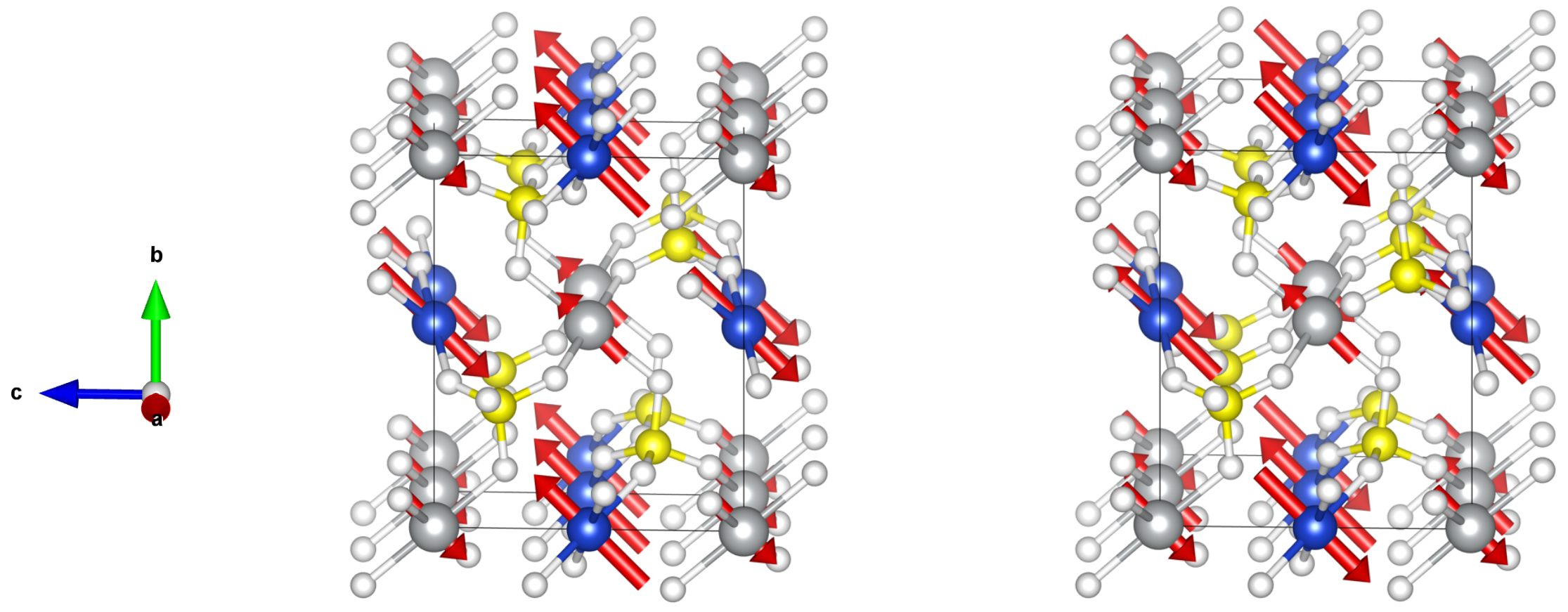

**Figure S13.** Spin orientation in AFM1 (energy groundstate using the GGA+U and HSE06 methods) and AFM5 (groundstate using SCAN method). The increased energy of ferromagnetic $J_1$ in SCAN calculations leads to the change of the groundstate to AFM5, where $J_{1D}$ is unfrustrated, while $J_{2D}$ is only partially frustrated. For the GGA+U and HSE06 methods, the interaction $J_1$ is weak and increases the energy of AFM1. The energies of various spin states and exchange couplings are provided in **Table S5** and **Table S6**.

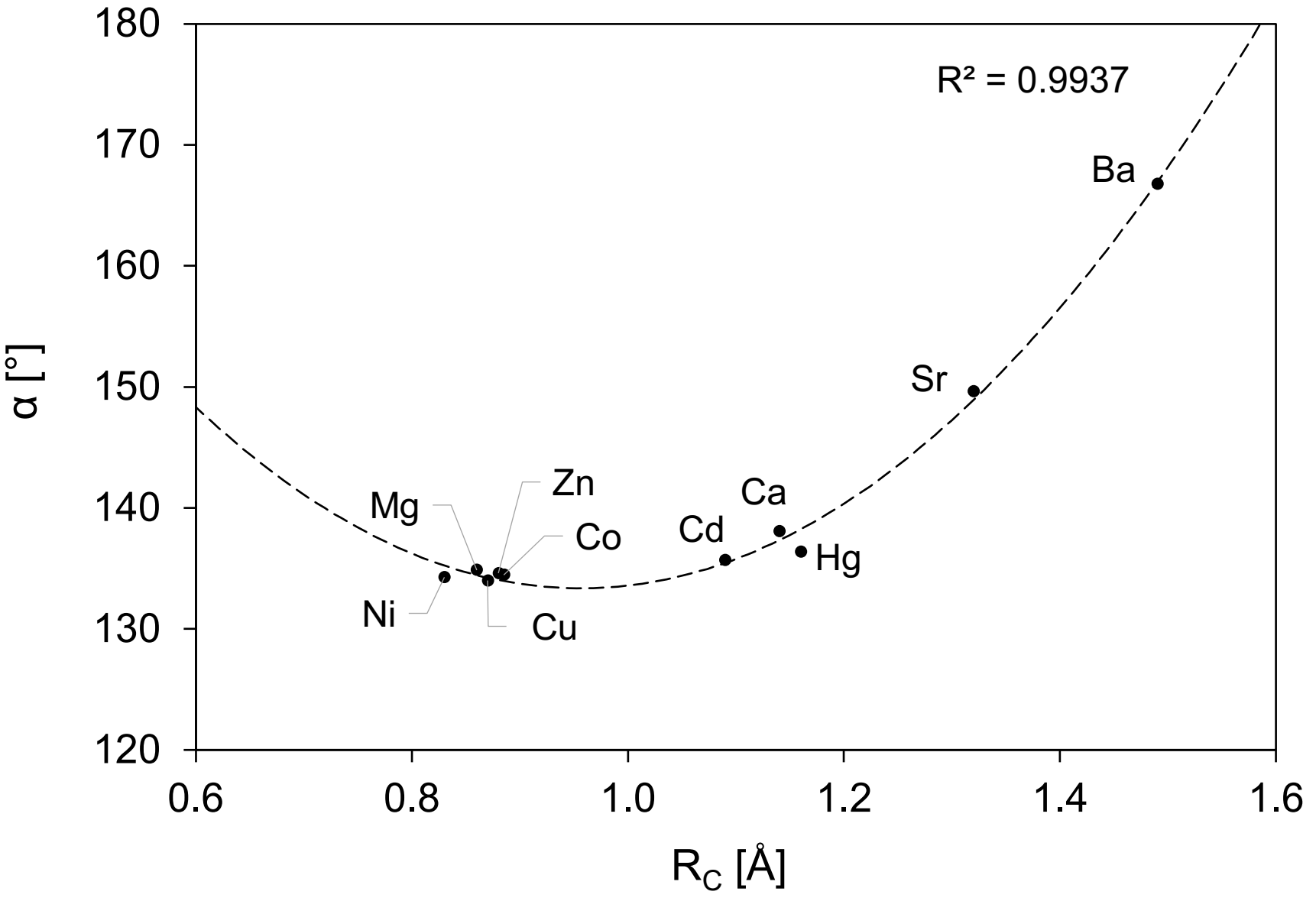

**Figure S14.** The relation between Ag'-S-Ag" intralayer angle (α) and $M^{2+}$ cation radius of the $MAg(SO_4)_2$ derivatives as optimized with the HSE06 method. The dashed line shows 2nd degree polynomial correlation.

**Table S7.** Geometric parameters of $[Ag(SO_4)_2]^{2-}$ layers in the $MAg(SO_4)_2$ derivatives as optimized with the HSE06 method. Bonds of Ag with O' and O" are intralayer, while bonds of Ag with O† are axial bonds of Ag-centered octahedra.

| Cation | $R_C$ [Å] | ΔE [kJ/mol] | $J_{2D}$ [meV] | α [°] | Ag'-Ag" [Å] | Ag-O' [Å] | Ag-O" [Å] | Ag···O†[Å] |
|---|---|---|---|---|---|---|---|---|
| $Zn^{2+}$ | 0.88 | –11.3 | –11.5 | 134.6 | 5.926 | 2.075 | 2.131 | 2.758 |
| $Cd^{2+}$ | 1.09 | –12.1 | –18.3 | 135.7 | 6.070 | 2.092 | 2.145 | 2.806 |
| $Hg^{2+}$ | 1.16 | 2.9 | –20.8 | 136.4 | 6.139 | 2.102 | 2.138 | 2.767 |
| $Mg^{2+}$ | 0.86 | –19.3 | –12.7 | 134.9 | 5.898 | 2.069 | 2.141 | 2.813 |
| $Ca^{2+}$ | 1.14 | 11.1 | –30.3 | 138.1 | 6.082 | 2.093 | 2.139 | 3.079 |
| $Sr^{2+}$ | 1.32 | 18.3 | –42.3 | 149.6 | 6.019 | 2.118 | 2.128 | 2.877 |
| $Ba^{2+}$ | 1.49 | 28.8 | –48.9 | 166.8 | 6.144 | 2.119 | 2.135 | 3.087 |
| $Cu^{2+}$ | 0.87 | –7.7 | –11.1 | 134.0 | 6.002 | 2.092 | 2.106 | 2.860 |
| $Ni^{2+}$ | 0.83 | –7.4 | –11.0 [a] | 134.3 | 5.871 | 2.069 | 2.126 | 2.753 |
| $Co^{2+}$ | 0.89 | –8.6 | –12.4 [a] | 134.5 | 5.913 | 2.075 | 2.132 | 2.753 |

[a] Indicated values of $J_{2D}$ are predicted from the logarithmic function fit.

## SI references

**CIF of $CuAg(SO_4)_2$**

```
#--------------------------------------------------------------------------
data__gsas2-md17
_audit_creation_method  "created in GSAS-II"

_pd_char_colour                 'brown'
_pd_calc_method                 'Rietveld Refinement'
_computing_structure_refinement
   "GSAS-II (Toby & Von Dreele, J. Appl. Cryst. 46, 544-549, 2013)"
_pd_proc_ls_prof_R_factor       0.01242
_pd_proc_ls_prof_wR_factor      0.01796
_pd_proc_ls_prof_wR_expected    0.01019
_refine_ls_matrix_type          full
_refine_ls_number_parameters    64
_refine_ls_goodness_of_fit_all  1.772
_refine_ls_shift/su_max         0.0266
_refine_diff_density_max        0.603
_refine_diff_density_min        -0.356
_refine_ls_wR_factor_obs        0.01798
_refine_ls_R_F_factor           0.02080
_refine_ls_R_Fsqd_factor        0.03205
_gsas_proc_ls_prof_R_B_factor   0.02156
_gsas_proc_ls_prof_wR_B_factor  0.03031

_diffrn_ambient_temperature  293
_diffrn_ambient_pressure     100
_diffrn_radiation_type  'Co K\a'
loop_
   _diffrn_radiation_wavelength
   _diffrn_radiation_wavelength_wt
   _diffrn_radiation_wavelength_id
  1.78892        1.0            1
  1.79278        0.5000         2

_chemical_name_common                  'copper(II) silver(II) sulfate(VI)'
_cell_length_a                         4.73365(8)
_cell_length_b                         8.71928(11)
_cell_length_c                         7.15754(16)
_cell_angle_alpha                      90.000000
_cell_angle_beta                       94.0852(9)
_cell_angle_gamma                      90.000000
_cell_volume                           294.670(9)
_space_group_name_H-M_alt              'P 21/n'
_space_group_IT_number                 14

_cell_formula_units_Z          2
_chemical_formula_sum          "Ag Cu O8 S2"
_chemical_formula_weight       363.53
_exptl_crystal_density_diffrn  4.0971

loop_
_space_group_symop_operation_xyz
   'x, y, z'
   '-x, -y, -z'
   '-x+1/2, y+1/2, -z+1/2'
   'x+1/2, -y+1/2, z+1/2'
```

```
loop_
   _atom_site_label
   _atom_site_type_symbol
   _atom_site_fract_x
   _atom_site_fract_y
   _atom_site_fract_z
   _atom_site_occupancy
   _atom_site_adp_type
   _atom_site_U_iso_or_equiv
   _atom_site_site_symmetry_multiplicity
Ag1    Ag   0.00000     0.00000     0.00000     1.0000     Uani 0.0240     2
Cu1    Cu   0.50000     0.50000     0.00000     1.0000     Uani 0.0240     2
S1     S    0.0211(6)   0.32308(24) 0.2086(4)   1.0000     Uiso 0.0200(8)  4
O1     O    0.0485(12)  0.1541(3)   0.2309(10)  1.0000     Uiso 0.0222(10) 4
O2     O    0.1379(11)  0.3873(5)   0.3880(6)   1.0000     Uiso 0.0222(10) 4
O3     O    -0.2784(8)  0.3732(6)   0.1841(8)   1.0000     Uiso 0.0222(10) 4
O4     O    0.1709(10)  0.3738(5)   0.0459(6)   1.0000     Uiso 0.0222(10) 4

loop_
   _atom_site_aniso_label
   _atom_site_aniso_U_11
   _atom_site_aniso_U_22
   _atom_site_aniso_U_33
   _atom_site_aniso_U_12
   _atom_site_aniso_U_13
   _atom_site_aniso_U_23
Ag1    0.0184(5)   0.0152(5)   0.0385(6)   -0.0066(21) 0.0040(7)   -0.0076(22)
Cu1    0.0184(5)   0.0152(5)   0.0385(6)   -0.0066(21) 0.0040(7)   -0.0076(22)
#----------------------------------------------------------------------
```

## DFT optimized crystal structures of $CuAg(SO_4)_2$ in POSCAR format

```
CuAg(SO4)2 P21/n optimized with GGA+U (GGA=PBEsol)
1.0
        4.7523102760         0.0000000000         0.0000000000
        0.0000000000         8.6793699265         0.0000000000
       -0.5552604754         0.0000000000         7.0971652899
   Ag   Cu    S    O
    2    2    4   16
Direct
     0.000000000         0.000000000         0.000000000
     0.500000000         0.500000000         0.500000000
     0.000000000         0.000000000         0.500000000
     0.500000000         0.500000000         0.000000000
     0.024282038         0.674128890         0.207728863
     0.975717962         0.325871050         0.792271137
     0.475502849         0.174054503         0.292402744
     0.524497151         0.825945497         0.707597256
     0.724694610         0.629542410         0.174679637
     0.275305450         0.370457590         0.825320363
     0.775267661         0.129746497         0.325476289
     0.224732339         0.870253563         0.674523711
     0.042479575         0.845836163         0.228595972
     0.957520485         0.154163897         0.771404028
     0.456719816         0.345738649         0.271390498
     0.543280184         0.654261291         0.728609502
     0.132718801         0.609580278         0.390147924
     0.867281258         0.390419722         0.609852135
     0.367293537         0.109275281         0.110059381
     0.632706404         0.890724719         0.889940560
     0.182082772         0.626800179         0.048379600
     0.817917168         0.373199761         0.951620400
     0.317919672         0.126710773         0.451807618
     0.682080328         0.873289227         0.548192382
```

```
CuAg(SO4)2 P21/n optimized with HSE06 (hybrid DFT)
1.0
        4.7321166992         0.0000000000         0.0000000000
        0.0000000000         8.6950702667         0.0000000000
       -0.5506966614         0.0000000000         7.1435090271
   Ag   Cu    S    O
    2    2    4   16
Direct
     0.000000000         0.000000000         0.000000000
     0.500000000         0.500000000         0.500000000
     0.000000000         0.000000000         0.500000000
     0.500000000         0.500000000         0.000000000
     0.023752630         0.674208760         0.206677005
     0.976247370         0.325791240         0.793322980
     0.476154625         0.174206376         0.293404043
     0.523845375         0.825793624         0.706595957
     0.726089954         0.629397154         0.176270932
     0.273910046         0.370602846         0.823729098
     0.773906589         0.129548192         0.323856950
```

```
     0.226093411         0.870451808         0.676143050
     0.040718138         0.843581438         0.224810690
     0.959281921         0.156418562         0.775189340
     0.458844364         0.343568802         0.275175571
     0.541155636         0.656431198         0.724824429
     0.134625375         0.610962152         0.385232687
     0.865374684         0.389037848         0.614767313
     0.365420401         0.110824943         0.114887297
     0.634579599         0.889175057         0.885112643
     0.178466082         0.626780033         0.049179841
     0.821533918         0.373219967         0.950820148
     0.321521044         0.126757801         0.450922638
     0.678478956         0.873242199         0.549077392

CuAg(SO4)2 P21/n optimized with SCAN (meta-GGA)
1.0
        4.7122426033         0.0000000000         0.0000000000
        0.0000000000         8.7800912857         0.0000000000
       -0.4218326703         0.0000000000         7.1340567972
   Ag   Cu    S    O
    2    2    4   16
Direct
     0.000000000         0.000000000         0.000000000
     0.500000000         0.500000000         0.500000000
     0.000000000         0.000000000         0.500000000
     0.500000000         0.500000000         0.000000000
     0.018999994         0.675585330         0.206795916
     0.981000006         0.324414670         0.793204129
     0.480935395         0.175586998         0.293263942
     0.519064546         0.824413002         0.706736028
     0.716870248         0.632942259         0.177512228
     0.283129752         0.367057741         0.822487772
     0.783115625         0.133031905         0.322587222
     0.216884375         0.866968155         0.677412808
     0.040457249         0.844200015         0.227267683
     0.959542751         0.155799985         0.772732317
     0.459281385         0.344192922         0.272705644
     0.540718555         0.655807018         0.727294385
     0.126598895         0.611020088         0.385724366
     0.873401165         0.388979912         0.614275634
     0.373441994         0.110918403         0.114376456
     0.626557946         0.889081597         0.885623515
     0.173693180         0.628007770         0.045040559
     0.826306760         0.371992290         0.954959393
     0.326281428         0.128028333         0.455041438
     0.673718572         0.871971607         0.544958591
```